\documentclass{pasa} 
\usepackage[figuresright]{rotating}
\usepackage{lscape}

\usepackage{graphics}

\usepackage{bigdelim}
\usepackage{bigstrut}

\usepackage{setspace}

\usepackage{natbib}
\usepackage{mathtools}
\usepackage{amsmath}
\usepackage{amssymb}
\usepackage{pifont}
\usepackage[subnum]{cases}

\newcommand{\placefigAsecond}{
  \begin{figure}
      \begin{center}
      \includegraphics[width=\columnwidth]{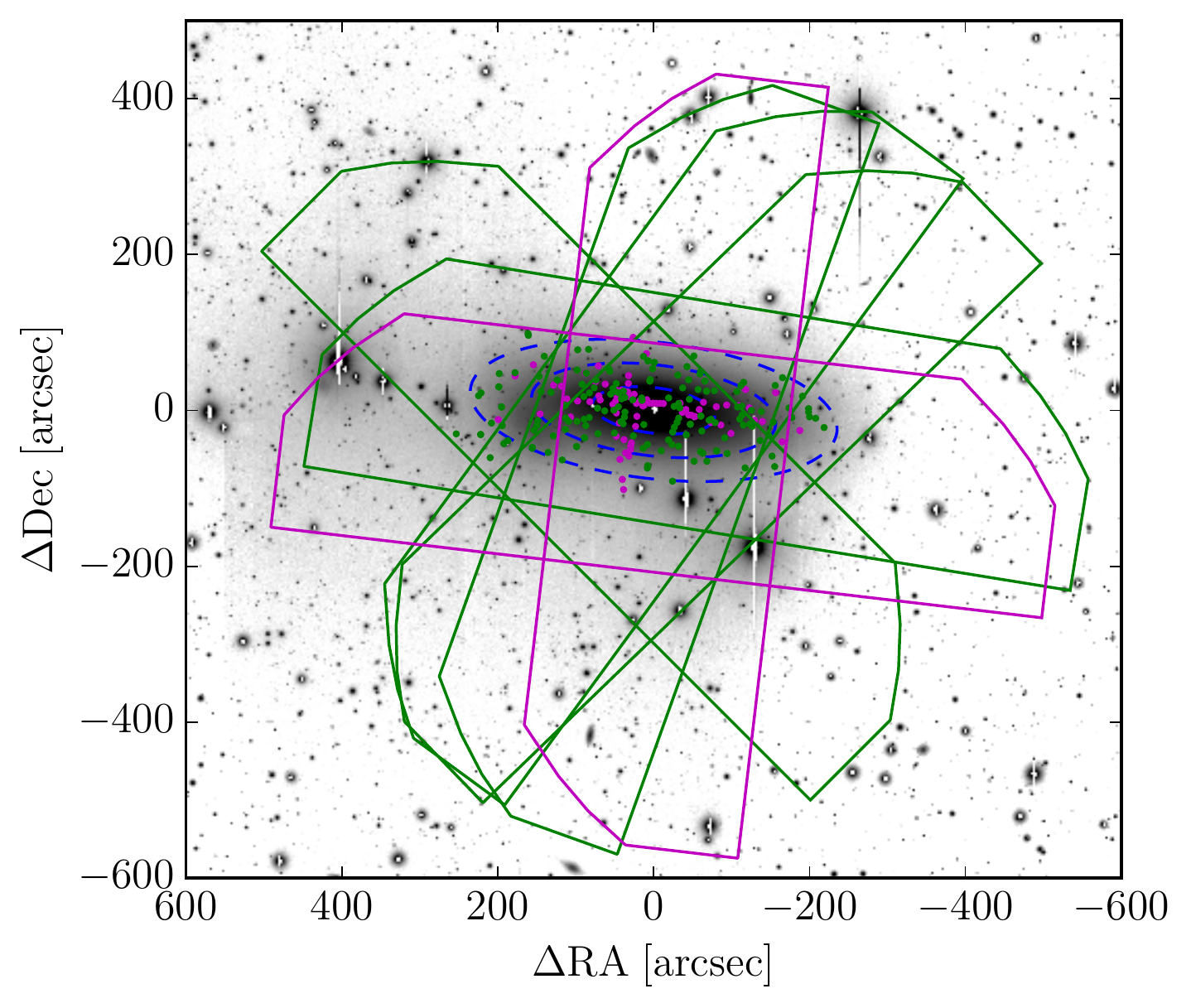}
    \end{center}
      \caption[]{
          Distribution of the observed 7 DEIMOS masks in the NGC~1023 field.
          The background image shows the NGC~1023 field observed in the \textit{r} filter with Subaru/Suprime-Cam.
          The green and magenta points show the positions of the spectra with S/N$>8$ for masks 1-5 and 6-7, respectively.
          The blue dashed lines present NGC~1023 $1$, $2$ and $3~\rm{R_{e}}$ isophotes.
          The footprints of the 7 DEIMOS masks are presented as solid lines.
          In particular, the magenta lines show the position of the 2 new \textit{SuperSKiMS} masks and the green lines the position of the 5 masks previously observed in the SLUGGS survey.
          Since the galaxy surface brightness decreases with distance from the centre, it is not possible to extract high S/N integrated stellar spectra from the outer slits.
          (This plot is best viewed in colour).
             }
    \label{fig:A}
  \end{figure}
}

\newcommand{\placefigB}{
  \begin{figure}
      \begin{center}
      \includegraphics[width=\columnwidth]{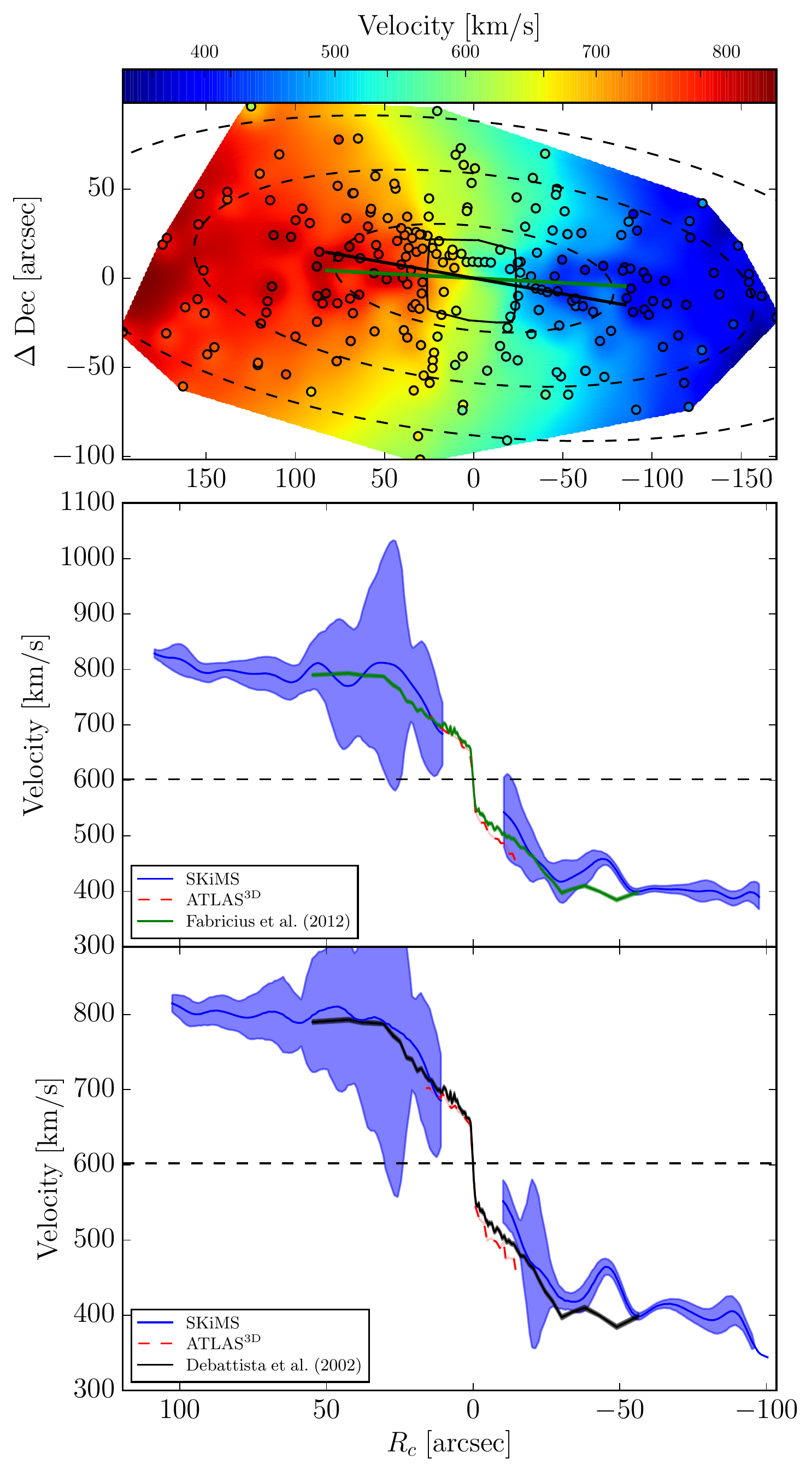}
    \end{center}
      \caption[]{
            NGC~1023 2D velocity map and 1D radial profiles.
            The \textit{top} panel presents the 2D stellar velocity map from kriging obtained from all our points.
            North is up and East is left.
            Both map pixels and points are colour coded according to their velocity as in the top colour bar.
            The black circles show the positions of the slits from which the kriged 2D map has been obtained.
            The $1$, $2$ and $3~\rm{R_{e}}$ isophotes are shown as dashed lines.
            The green and black lines show, respectively, the position of \citet{Fabricius12} and \citet{Debattista02} major axis slits.
            The solid black line show the field-of-view of \atlas.
            In the \textit{middle} and \textit{bottom} panels we present the velocity radial profiles from literature longslits compared with those extracted from a virtual longslit with the same $PA$ and width.
            In both panels, the profiles extracted from our kriging map and the \atlas\ map are plotted as solid blue and red dashed lines, respectively.
            The light blue contour shows the $1\sigma$ confidence region for our profile.
            We compare these profiles with the longslit data of \citet[\textit{middle} panel, green line]{Fabricius12} and \citet[\textit{bottom} panel, black line]{Debattista02}.
            The black dashed line shows the systemic velocity of NGC~1023.
            In general, the virtual slits show good agreement with the literature longslits in the overlapping regions at positive radii (Eastern side), while at negative radii the agreement is weaker.
             }
    \label{fig:B}
  \end{figure}
}

\newcommand{\placefigC}{
  \begin{figure}
      \begin{center}
      \includegraphics[width=\columnwidth]{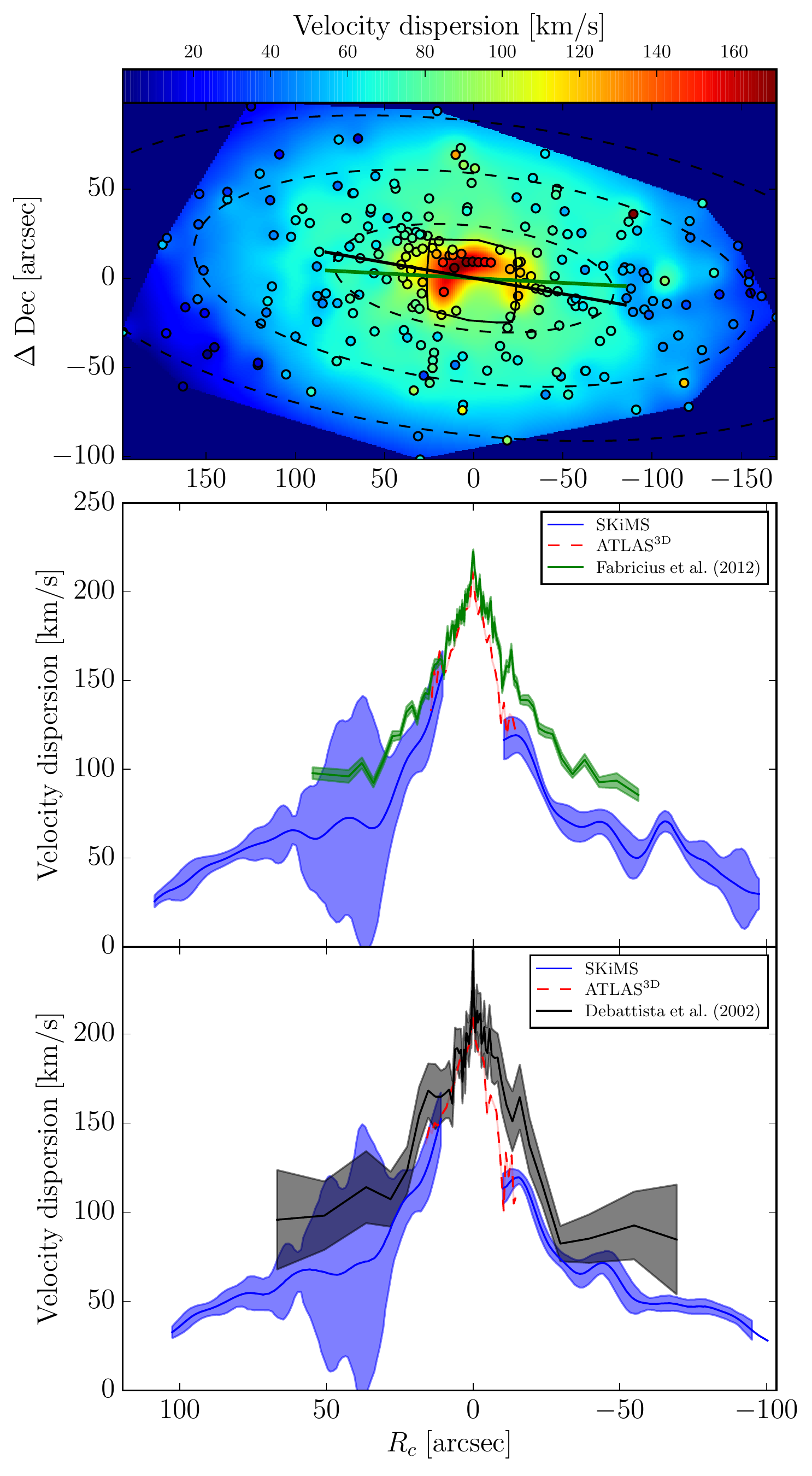}
    \end{center}
      \caption[]{
            As Figure \ref{fig:C}, but for the velocity dispersion.
            In the overlapping regions, the velocity dispersion profiles of both \atlas\ and our kriging virtual slits show a remarkable agreement.
            Both the \citet{Fabricius12} and \citet{Debattista02} show higher velocity dispersions at large radii than the virtual slit profiles.
            This may be a consequence of their increasingly larger spatial bins with radius, which can artificially increase the measured velocity dispersion.
            }
    \label{fig:C}
  \end{figure}
}

\newcommand{\placefigD}{
  \begin{figure}
      \begin{center}
      \includegraphics[width=\columnwidth]{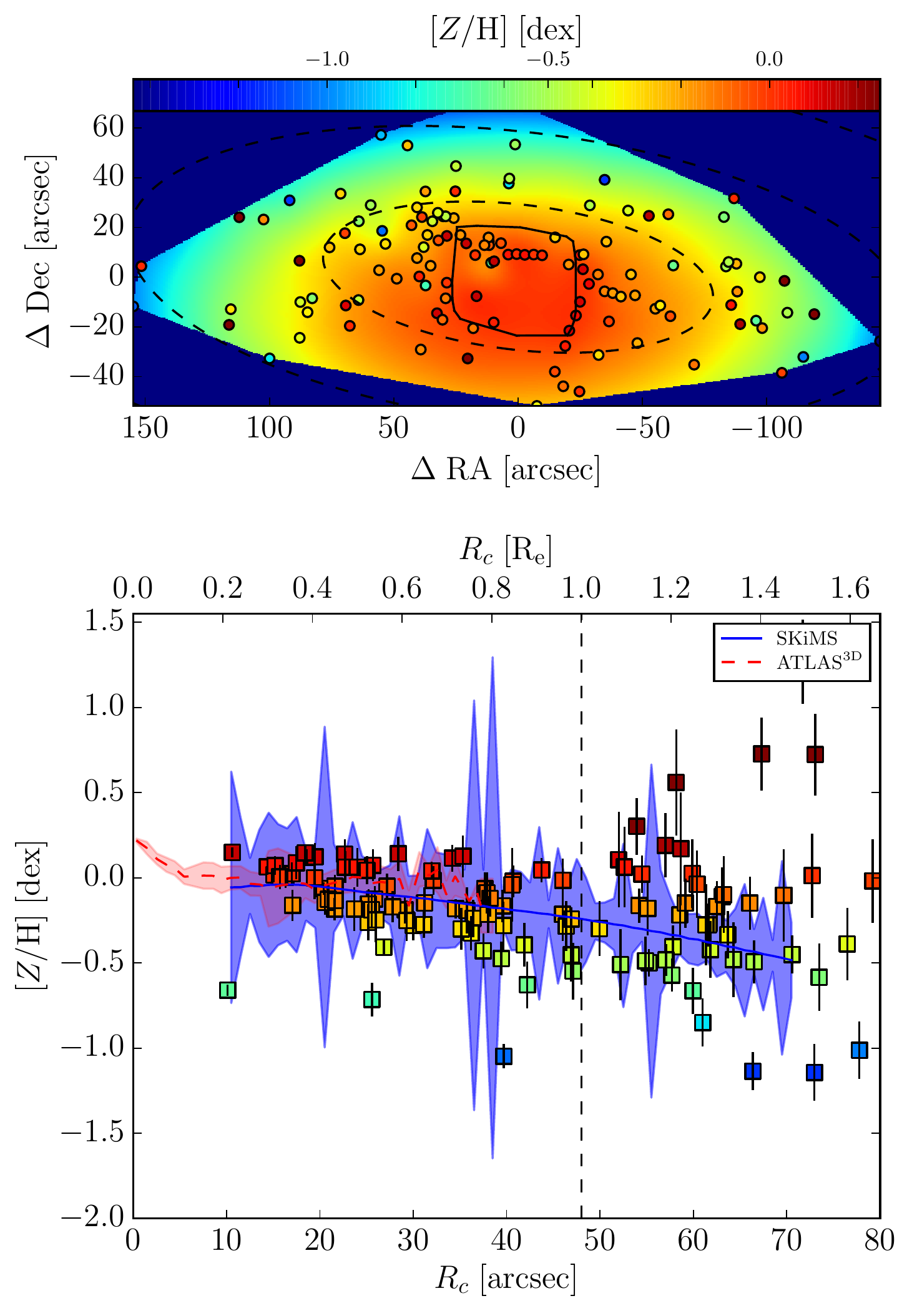}
    \end{center}
      \caption[]{
            NGC~1023 2D metallicity map and 1D azimuthally-averaged radial profiles.
            The \textit{top} panel shows the 2D stellar metallicity map from kriging obtained from all our points (i.e. masks 1 to 7).
            North is up and East is left.
            Both map pixels and datapoints are colour coded according to their metallicity as in the top colour bar.
            The $1$, $2$ and $3~\rm{R_{e}}$ isophotes are shown as dashed lines.
            The solid black line show the field-of-view of \atlas.
            In the \textit{bottom} panel we plot the azimuthally averaged metallicity profiles against the circularised galactocentric radius measured from the \atlas\ and our kriged maps as dashed red and solid blue lines, respectively.
            The inner limit of our radial profile is given by the position of the innermost available slit.
            The vertical dashed line shows the scale of $1~\rm{R_{e}}$.
            In the same plot we show the measured metallicity datapoints as square symbols, colour coded according to the colour scale at the top.
             }
    \label{fig:D}
  \end{figure}
}

\newcommand{\placefigF}{
  \begin{figure}
      \begin{center}
      \includegraphics[width=\columnwidth]{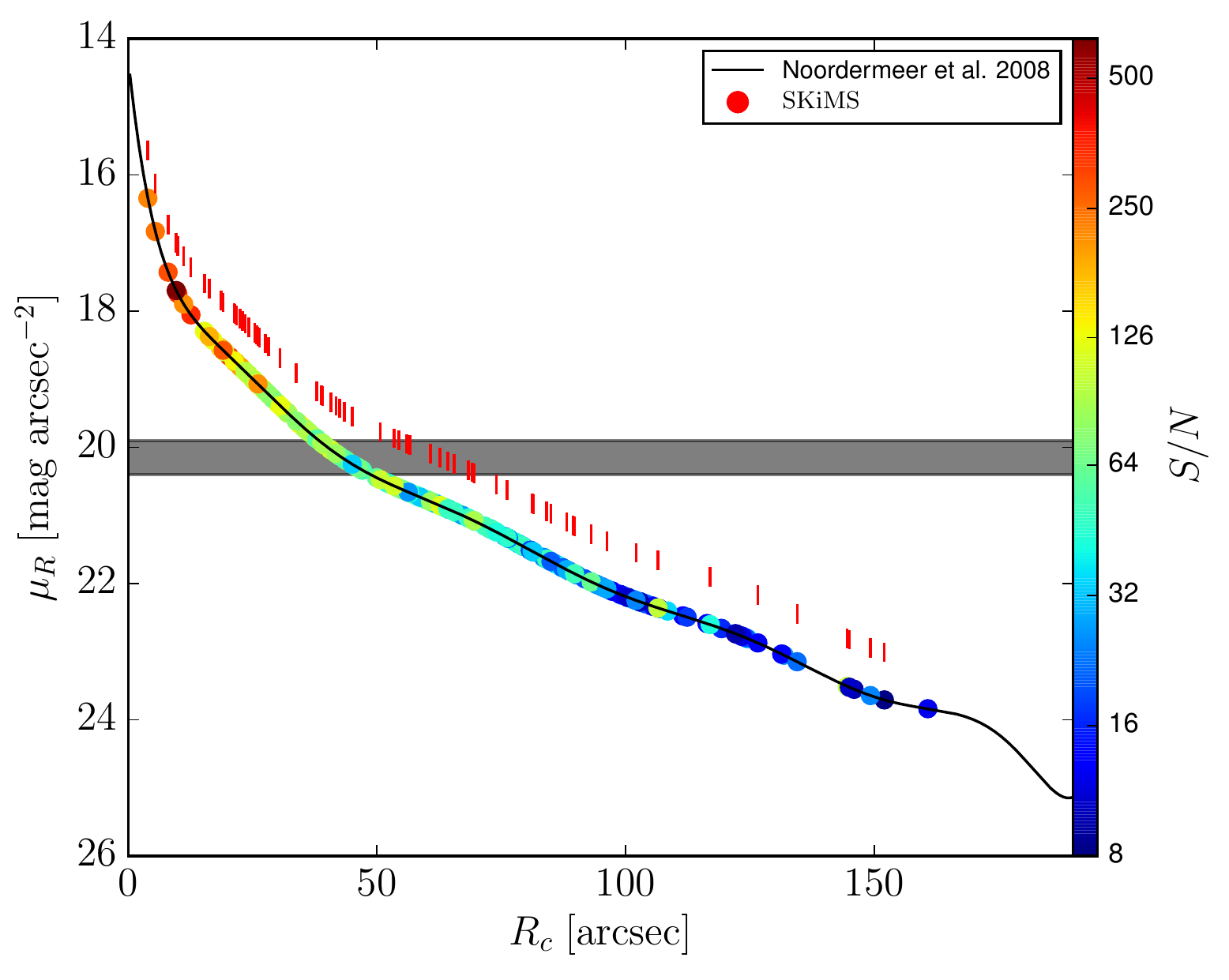}
    \end{center}
      \caption[]{
            NGC~1023 surface brightness profile and spectra's S/N radial distribution.
            The black line shows the NGC~1023 surface brightness profile in R-band from \citet{Noordermeer08} against the circularised galactocentric radius.
            The circles show the positions of the spectra obtained from SKiMS with S/N$>8$ in all the 7 DEIMOS masks.
            All the circles are colour coded according to their S/N values, as in the colour bar on the right hand side.
            The red lines show the position of the SKiMS points from masks 6 and 7 (vertically offset for a better visualisation).
            The gray horizontal shade shows the typical average sky R-band surface brightness at Mauna Kea in dark time.
            In particular, within $30~\rm{arcsec}$ from the galaxy centre spectra have generally S/N$>100$, while at $1~\rm{R_{e}}$ (i.e.  S/N$\approx40$).
            We are able to obtain the stellar kinematics (i.e. S/N$>8$) out to more than $160$ arcsec, where the surface brightness of the target is several magnitudes fainter than the sky.
            (This plot is best viewed in colour).
             }
    \label{fig:F}
  \end{figure}
}

\newcommand{\placefigH}{
  \begin{figure*}
      \begin{center}
      \includegraphics[width=2\columnwidth]{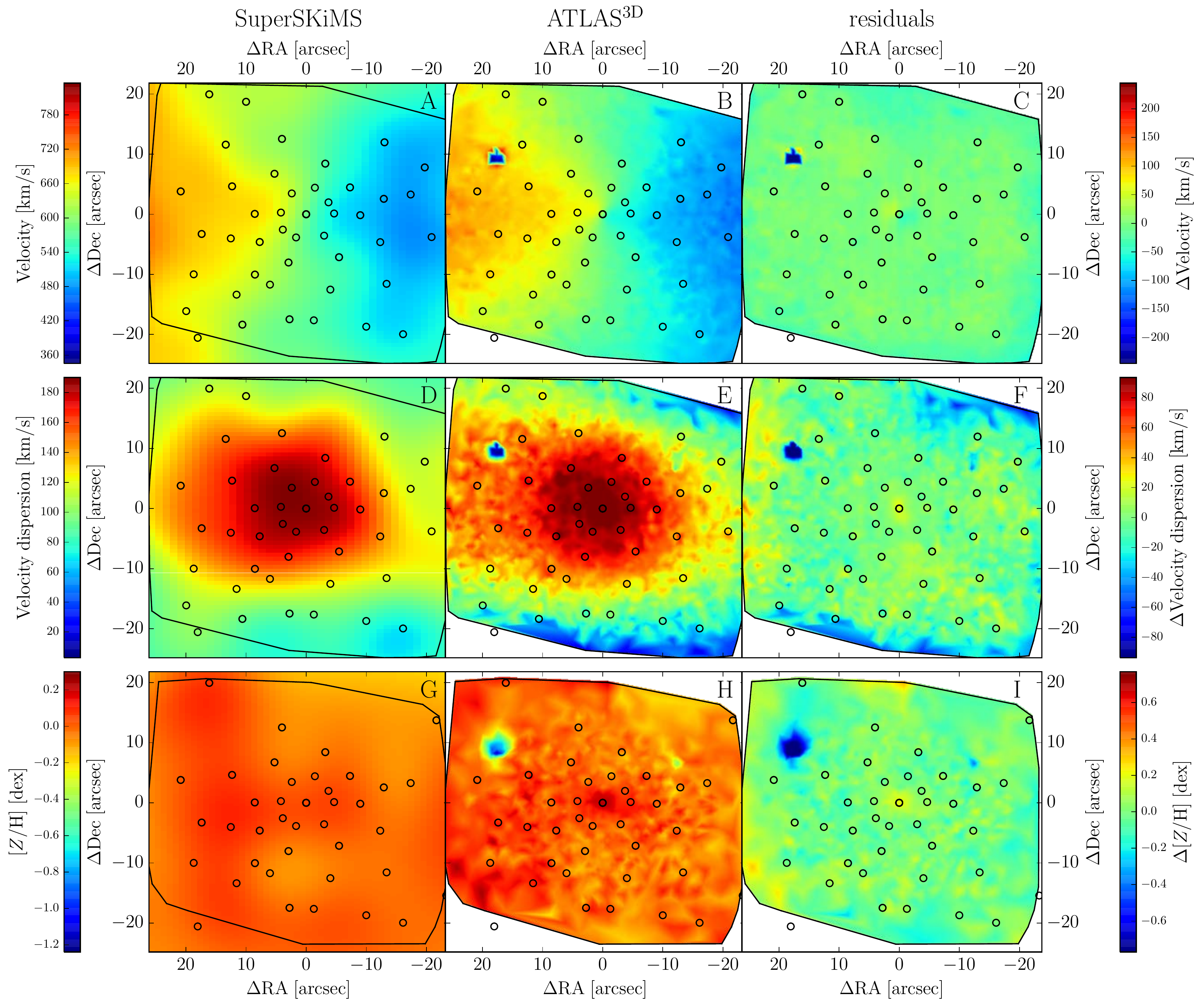}
    \end{center}
      \caption[]{
            Comparisons between \textit{SuperSKiMS} and \atlas\ 2D maps.
            From \textit{top} to \textit{bottom}, the first column of panels shows the 2D velocity (A), velocity dispersion (D) and $[Z/$H] (G) kriged maps for the central regions of NGC~1023.
            These maps are obtained with kriging from a mocked \textit{SuperSKiMS} dataset, in which the slits in the \atlas\ field-of-view are extracted from the \atlas\ kinematics and metallicity maps.
            Panels B, E and H show the \atlas\ velocity, velocity dispersion and metallicity 2D maps, respectively, on the same spatial scale.
            Since \atlas\ data is spatially binned, here we apply linear interpolation in order to obtain smooth 2D maps.
            In the third column of panels (C, F and I) the differences between the \atlas\ and the \textit{SuperSKiMS} inner region maps are shown, colour coded according to the colour bars on the right hand side.
            Since the difference is measured on the \atlas\ datapoint locations, we apply linear interpolation in order to obtain smooth 2D maps.
            The right side colour bars have the same dynamical range as those on the left side of the plot.
            In all the maps, the black circles show the positions of the \textit{SuperSKiMS} slits and the solid line shows the layout of the \atlas\ data.
            Note that the Galactic star visible in the \atlas\ field in the top-left corner is not present in the \textit{SuperSKiMS} maps, since no slits have been placed on it.
             }
    \label{fig:H}
  \end{figure*}
}

\newcommand{\placefigI}{
  \begin{figure*}
      \begin{center}
      \includegraphics[width=2\columnwidth]{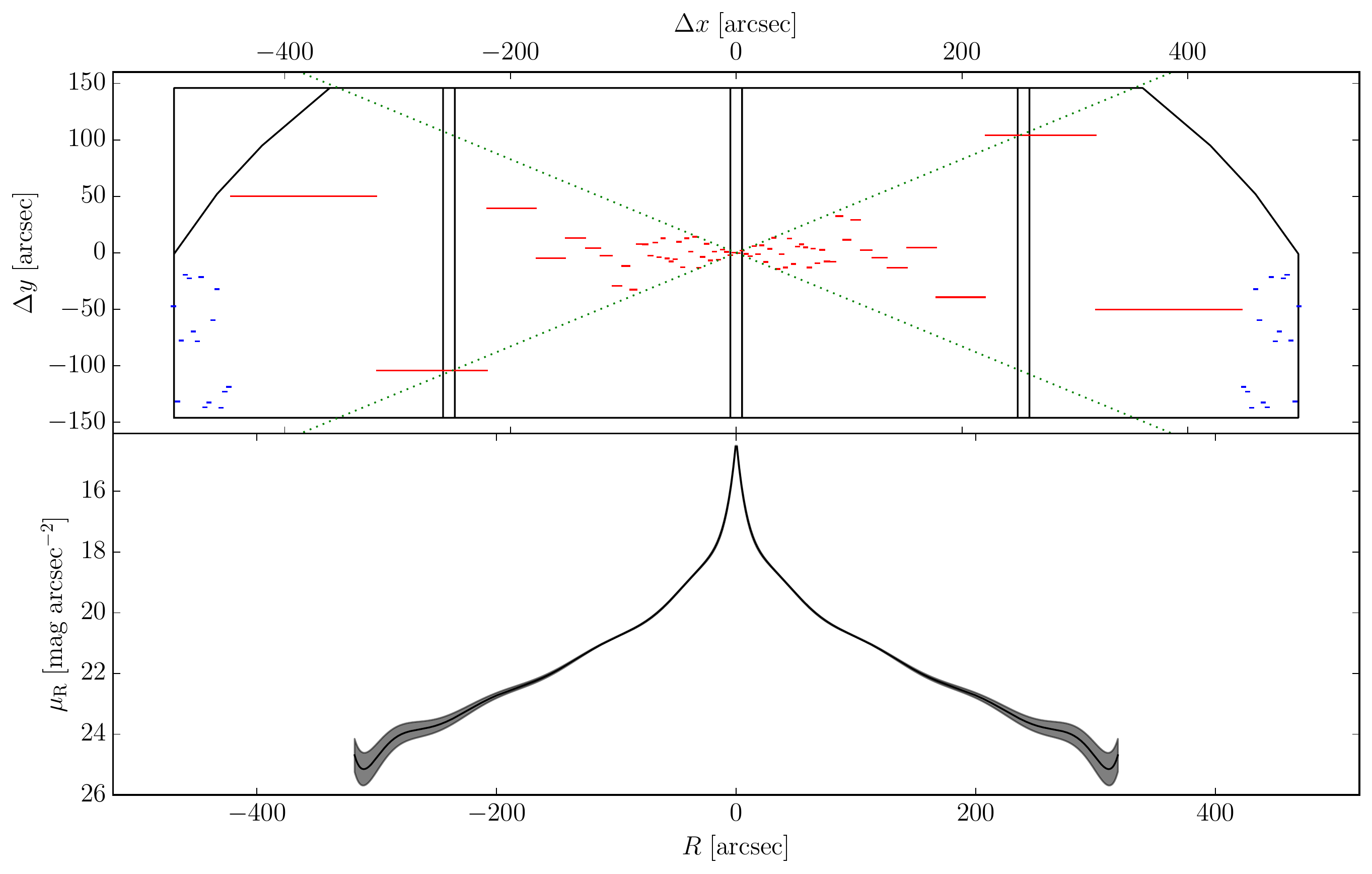}
    \end{center}
      \caption[]{
            \textit{SuperSKiMS} slit distribution for a NGC~1023 mask assuming 4 slit mask pointings.
            The \textit{upper} panel shows the layout of a DEIMOS mask optimised for the best azimuthal coverage.
            The red lines show the positions and lengths of the slits targeting the integrated stellar population.
            The blue lines show the positions of the ``sky'' slits used for the non-local sky subtraction.
            The green dotted lines show the two 45 degrees cones within which the slits are distributed.
            In this configuration, four masks are required to cover the whole azimuthal range (i.e. 360 degrees).
            In the \textit{bottom} panel the NGC~1023 R-band photometric profile along the major axis by \citet{Noordermeer08} is plotted as a black solid line.
            In this case, the $x$ axis represents the real galactocentric radius along the major axis.
             }
    \label{fig:I}
  \end{figure*}
}

\newcommand{\placefigJ}{
  \begin{figure*}
      \begin{center}
      \includegraphics[width=2\columnwidth]{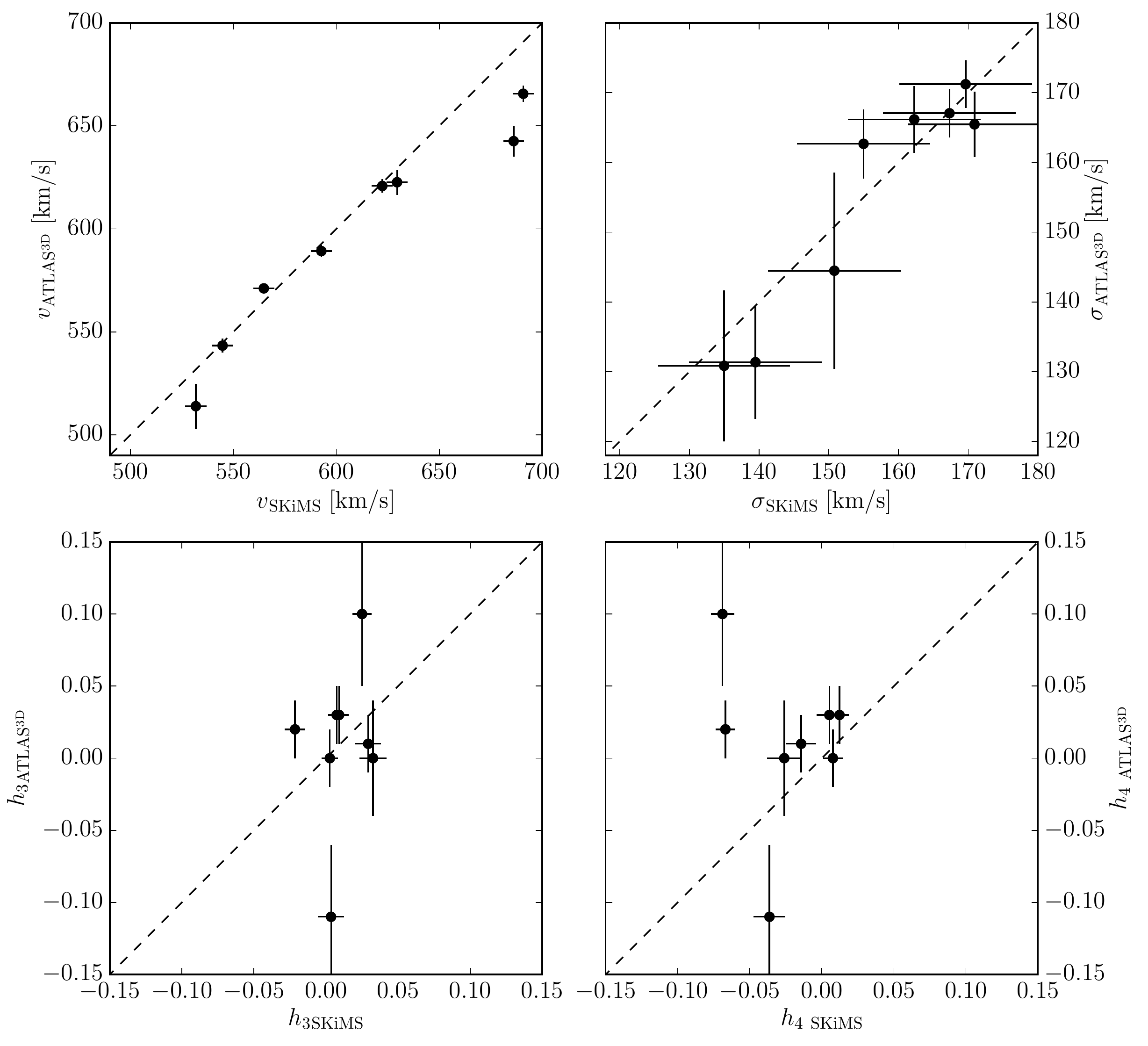}
    \end{center}
      \caption[]{
            Comparison between SKiMS (masks 1 to 7) and \atlas\ kinematic data points.
            The \textit{top~left} panel shows on the x axis the velocities obtained from the SKiMS points within the \atlas\ field-of-view.
            On the y axis the \atlas\ velocities at the same spatial locations are shown.
            The dashed line shows the one-to-one relation.
            Overall the two datasets show a good agreement.
            In the \textit{top~right} panel the velocity dispersion measurements of the two datasets are compared.
            Again, the dashed line present the one-to-one relation.
            We do not observe any significant offset between the velocity dispersions of the two datasets.
            The \textit{bottom~left} and \textit{bottom~right} panels show the comparison between the $h_{3}$ and $h_{4}$ values of the two datasets, respectively.
            Despite the narrow dynamical range covered, a reasonable agreement between \atlas\ and SKiMS values is seen in both $h_{3}$ and $h_{4}$.
             }
    \label{fig:J}
  \end{figure*}
}

\newcommand{\placefigK}{
  \begin{figure*}
      \begin{center}
      \includegraphics[width=2\columnwidth]{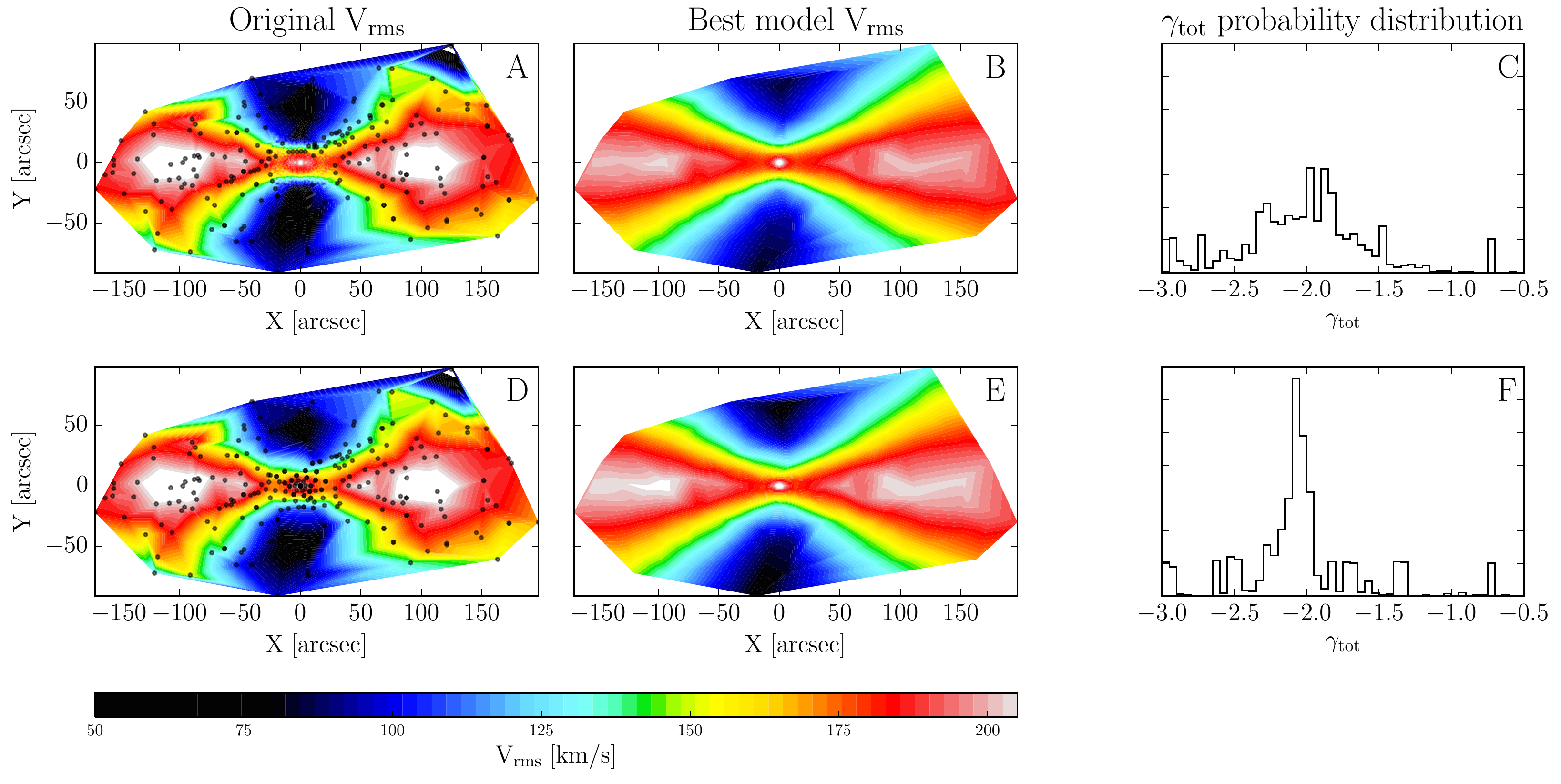}
    \end{center}
      \caption[]{
              Comparison between the $V_{\rm{rms}}$ input data and best fit model obtained from the two different datasets.
              Panels A and D show the $V_{\rm{rms}}$ 2D distribution obtained from the SLUGGS+\atlas\ and \textit{SuperSKiMS} datasets, respectively.
              The black points show the position of the SLUGGS datapoints and the mocked \textit{SuperSKiMS} datapoints in Panels A and D, respectively.
              The best-fit JAM models in the two cases are displayed in panels B and E.
              All the $V_{\rm{rms}}$ maps are colour coded according to the colourbar.
              On the right hand side of the plot, panels C and D show the posterior total mass slope $\gamma_{\rm{tot}}$ probability distribution in the SLUGGS+\atlas\ and \textit{SuperSKiMS} cases, respectively.
              The two density distributions peak at similar values of $\gamma_{\rm{tot}}$.
              }
    \label{fig:K}
  \end{figure*}
}

\newcommand{\placefigL}{
  \begin{figure}
      \begin{center}
      \includegraphics[width=\columnwidth]{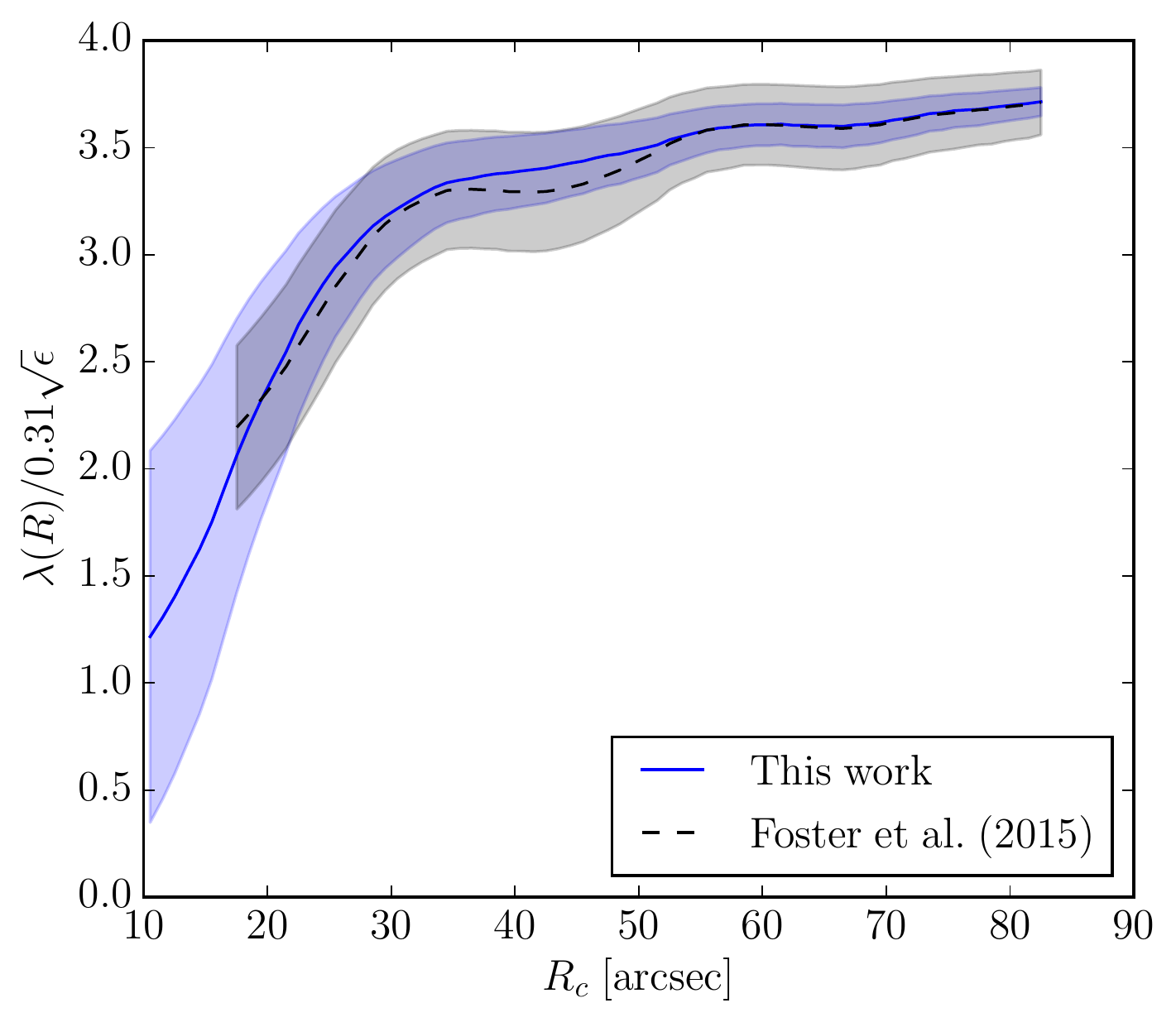}
    \end{center}
      \caption[]{
          Radial profile of the local specific angular momentum in NGC~1023.
          The solid blue line shows the profile measured from the kinematic data presented in this paper.
          The dashed black line is the local specific angular momentum profile as extracted from the kinematic data presented in \citet{Foster15}.
          In the first case, the higher number of datapoints (particularly in the inner regions) better probes the galaxy inner regions and slightly reduces the profile uncertainties.
          }
    \label{fig:L}
  \end{figure}
}

\newcommand{\placetabA}
{ \begin{table}
  \begin{center}
\resizebox{\columnwidth}{!}
{
  \begin{tabular}{cccccc}
     \hline

      Mask number & $PA$  & Date    & Data  & Exp. time   & Seeing  \\
     & (degrees)      &     &             & (seconds)     & (arcsec)  \\
     \hline
    $1$ & $143.8$   & 2011-11-30 & $39$ & $5398$ & $1.1$ \\ 
    $2$ & $80.8$    & 2011-11-30 & $22$ & $5399$ & $1.1$ \\ 
    $3$ & $45.0$    & 2012-01-16 & $34$ & $8839$ & $1.3$ \\ 
    $4$ & $134.0$   & 2012-01-17 & $28$ & $4920$ & $1.4$ \\ 
    $5$ & $160.3$   & 2013-09-29 & $41$ & $7200$ & $0.8$ \\ 
    \hline
    $6$ & $83.3$    & 2014-10-27 & $51$ & $10800$ & $1.1$ \\
    $7$ & $173.3$   & 2014-10-27 & $23$ & $3000$ &  $1.3$ \\ 
    \hline
  \end{tabular}
}
  \end{center}
    \caption{DEIMOS mask parameters.
    The columns present: (1) mask number, (2) mask orientation, (3) observation date, (4) number of extracted spectra with S/N$>8$, (5) mask exposure time and (6) average seeing during the observation.}
  \label{tab:A}
  \end{table}
}

\newcommand{\placetabC}
{ \begin{table}
  \begin{center}
\resizebox{\columnwidth}{!}{%
  \begin{tabular}{ccc}
     \hline

        & Inner & Outer \\
        & $\rm{(dex~dex^{-1})}$ & $\rm{(dex~dex^{-1})}$ \\
  \hline
  Forbes et al. (2016) & $-0.28^{+0.12}_{-0.13}$ &  $-0.94^{+0.49}_{-0.65}$\\
  This work          & $-0.38^{+1.19}_{-1.01}$ &  $-1.43^{+0.63}_{-0.60}$\\
    \hline
  \end{tabular}
}
  \end{center}
    \caption{Comparison between old and new stellar metallicity gradients of NGC~1023. }
  \label{tab:C}
  \end{table}
}

\newcommand{\sauron}{SAURON}
\newcommand{\atlas}{ATLAS$^{\rm{3D}}$}

\title[SuperSKiMS]{The SLUGGS survey: a new mask design to reconstruct the stellar populations and kinematics of both inner and outer galaxy regions}

\author[Pastorello et al.]{´Nicola~Pastorello$^{1,2}$\thanks{email: nicola.pastorello@deakin.edu.au}, Duncan~A.~Forbes$^{2}$, Adriano~Poci$^{3}$, Aaron~J.~Romanowsky$^{4,5}$, Richard~McDermid$^{3,6}$, Adebusola~B.~Alabi$^{2}$, Jean~P.~Brodie$^{5}$, Michele~Cappellari$^{7}$, Vincenzo~Pota$^{2,8}$, Caroline~Foster$^{6}$\\
\affil{$^1$Deakin Software and Technology Innovation Laboratory, Deakin University, Burwood, VIC 3125, Australia}
\affil{$^2$Centre for Astrophysics \& Supercomputing, Swinburne University, Hawthorn VIC 3122, Australia}
\affil{$^3$Department of Physics and Astronomy, Faculty of Science and Engineering, Macquarie University, Sydney, NSW 2109, Australia}
\affil{$^4$Department of Physics and Astronomy, San Jos\'e State University, One Washington Square, San Jos\'e, CA 95192, USA}
\affil{$^5$University of California Observatories, 1156 High Street, Santa Cruz, CA 95064, USA}
\affil{$^6$Australian Astronomical Observatory, PO Box 915, North Ryde, NSW 1670, Australia}
\affil{$^7$Sub-department of Astrophysics, Department of Physics, University of Oxford, Denys Wilkinson Building, Keble Road, Oxford, \\OX1 3RH, UK}
\affil{$^8$INAF - Osservatorio Astronomico di Capodimonte, Salita Moiariello, 16, I-80131 Napoli, Italy}
}

\begin{document}

\begin{abstract}
  Integral field unit spectrographs allow the 2D exploration of the kinematics and stellar populations of galaxies, although they are generally restricted to small fields-of-view.
  Using the large field-of-view of the DEIMOS multislit spectrograph on Keck and our Stellar Kinematics using Multiple Slits (SKiMS) technique, we are able to extract sky-subtracted stellar light spectra to large galactocentric radii.
  Here we present a new DEIMOS mask design named \textit{SuperSKiMS} that explores large spatial scales without sacrificing high spatial sampling.
  We simulate a set of observations with such a mask design on the nearby galaxy NGC~1023, showing that the kinematic and metallicity measurements can reach radii where the galaxy surface brightness is several orders of magnitude fainter than the sky.
  Such a technique is also able to reproduce the kinematic and metallicity 2D distributions obtained from literature integral field spectroscopy in the innermost galaxy regions.
  In particular, we use the simulated NGC~1023 kinematics to model its total mass distribution to large radii, obtaining comparable results with those from published integral field unit observation.
  Finally, from new spectra of NGC~1023 we obtain stellar 2D kinematics and metallicity distributions that show good agreement with integral field spectroscopy results in the overlapping regions.
  In particular, we do not find a significant offset between our SKiMS and the \atlas\ stellar velocity dispersion at the same spatial locations.

\end{abstract}


\maketitle


\section{Introduction}
\label{sec:introduction}
  In recent years the two-dimensional (2D) distribution of galaxy properties has provided a remarkably high number of useful constraints to understand galaxy formation and evolution processes.
  For example, the stellar and gas 2D line-of-sight kinematics are strongly linked with galaxy intrinsic shape, internal orbital structure and radial mass-to-light ratio ($M/L$) profile (e.g. \citealt{vanderMarel93, Gerhard93, Cretton00,Cappellari13b}).
  Furthermore, from absorption line index and/or full spectral fitting analyses it is possible to extract the luminosity- and mass-weighted parameters of the integrated stellar population (e.g. age, total metallicity $[Z/\rm{H}]$, $\alpha$-element abundance $[\alpha/\rm{Fe}]$, stellar initial mass function), as well as single chemical element abundances
  (e.g. \citealt{Kuntschner10, McDermid15}).

  Long-slit spectroscopy, along different position angles in a galaxy, has been used in the past to access this 2D information (e.g. \citealt{Davies88,Statler99, Saglia10}).
  Even though this approach provides information out to large radii, it is incapable of properly mapping the galaxy internal structure and requires a large amount of telescope time \citep{Statler94, Arnold94}.
  Furthermore, the spectra are obtained at different times, suffering from systematic effects in the case of imperfect sky subtraction.

  A way to overcome such issues is the use of Integral Field Unit (IFU) spectrographs.
  These instruments are able to obtain full spectral coverage of a 2D field-of-view (FoV) with a single exposure.
  Because of their efficiency, IFU spectrographs have been extensively used in surveys to explore the properties of large numbers of galaxies in the nearby Universe, although without being able to explore out to more than a few effective radii ($\rm{R_{e}}$).

  For instance, the \atlas\ survey used the \sauron\ IFU \citep{Bacon01} to explore the kinematics and the stellar populations of 260 local early-type galaxies (ETGs) in their inner regions (i.e. probing $R<1~\rm{R_{e}}$ in most cases, \citealt{Cappellari11a}).
  Similarly, the Sydney-AAO Multiobject Integral-field spectrograph (SAMI) is used in the SAMI survey to observe $3400$ galaxies, reaching not much beyond $2~\rm{R_{e}}$ \citep{Bryant15}.

  In some cases, IFU spectrographs have been used to explore the galaxy chemodynamics beyond $1~\rm{R_{e}}$.
  For example, using the SAURON spectrograph, \citet{Weijmans09} measured both kinematics and absorption line-strengths in NGC~821 and NGC~3379 out to almost $4~\rm{R_{e}}$.
  With a larger sample, the Visible Integral-field Replicable Unit Spectrograph prototype (VIRUS-P) has been used to study the kinematics \citep{Raskutti14} and stellar populations parameters \citep{Greene13} of ETGs out to $R\approx2.5~\rm{R_{e}}$, although with poor velocity resolution (i.e. $\sigma\approx150~\rm{km~s^{-1}}$) and low S/N ratio at large radii.
  The same instrument is also used for the MASSIVE survey, which targets the most massive ETGs (i.e. $M_\star \geq10^{11.5}~\rm{M_{\odot}}$, \citealt{Ma14}).
  Furthermore, the Calar Alto Legacy Integral Field Area Survey (CALIFA) takes advantage of the PMAS/PPAK instrument and aims to observe 600 galaxies in the local Universe out to generally $2~\rm{R_{e}}$ \citep{Sanchez12}.
  Finally, an even larger sample of galaxies is that explored by the Mapping Nearby Galaxies at APO (MaNGA, i.e. $\approx10000$ galaxies), but with a field-of-view still limited to the inner $1.5-2.5~\rm{R_{e}}$ \citep{Bundy15}.

  As shown by \citet{Brodie14}, in a typical ETG only a small fraction of the total galaxy mass and angular momentum is included in the inner few $\rm{R_{e}}$.
  Furthermore, while these inner regions are dominated in mass by the stellar component, at large radii the dominant component is the dark matter.
  In order to sample a sufficiently large fraction of total galaxy mass and angular momentum, good 2D spatial coverage is needed in both these regions.
  In particular, it is important to consistently measure the stellar kinematics and population parameters at both small and large radii.
  Applying the Stellar Kinematics from Multiple Slits (SKiMS) method \citep{Norris08, Proctor09, Foster09}, it is possible to measure both the kinematics and the metallicity of nearby galaxy stellar components out to large radii (e.g. \citealt{Foster13, Arnold14, Pastorello14}).
  In particular, \citet{Foster15} obtained SKiMS kinematic measurements out to $5~\rm{R_{e}}$ ($2.6~\rm{R_{e}}$ on average) and \citet{Pastorello14} metallicity measurements out to about $3~\rm{R_{e}}$.
  The original SKiMS method was developed to extract the background galaxy stellar light spectra from the same DEIMOS slits that were primarily targeting globular cluster (GC) candidates.
  In this way, the galaxy stellar component is probed at randomly scattered spatial positions.
  Furthermore, the bright inner galaxy regions are not targeted, since GCs are difficult to detect in the presence of a strong stellar background.
  In order to include the innermost regions in their dynamical models of 14 ETGs, \citet{Cappellari15} combined the SKiMS large-radii stellar kinematics with those from \atlas\ at small radii.
  The use of two different datasets required the accounting for a number of systematic issues given by the non-homogeneity of the data (e.g. different spatial sampling and kinematic uncertainties).
  A way to overcome such issues is to use an homogeneous, although optimal azimuthally and spatially distributed, kinematic dataset.

  Here we present a new mask design that takes advantage of the Keck/DEIMOS multislit spectrograph to reliably explore the stellar kinematics and stellar population parameters in ETGs out to large radii, with complete azimuthal coverage.
  We name this \textit{SuperSKiMS}, since it is an application of the SKiMS method on data obtained using specially designed multislit masks.
  In particular, the Keck/DEIMOS slit distribution in the \textit{SuperSKiMS} configuration allows for the optimal sampling of both the inner and the outer regions of nearby galaxies.
  Tho demonstrate this, we present the results from mock \textit{SuperSKiMS} simulations in order to show that such a technique can return data comparable to that from IFU spectroscopy.
  In particular, we use such simulated observations to extract the total mass profile slope of the nearby lenticular galaxy NGC~1023 from the modelling of the galaxy kinematics.
  These results are compared with those obtained from \atlas\ kinematic measurements in the centre and with the results by \citet{Cappellari15} (i.e. using a similar but smaller sample, with a shorter radial baseline).

  Moreover, we use two SuperSKiMS masks to obtain new stellar kinematics and metallicity measurements for NGC~1023.
  These values are used together with already published SKiMS measurements, thus extending the study of stellar kinematics and metallicity out to $3.6$ and $2.5~\rm{R_{e}}$, respectively.
  At such galactocentric radii the stellar light from the galaxy is just 1.6\% and 5.5\% of the sky flux at similar wavelengths.
  At the same time, our new data extend into the central regions of NGC~1023, overlapping with several published longslit and IFU observations.

  We then evaluate how the addition of these new datapoints to the dataset affects several results already presented in the SLUGGS survey.
  In particular, from the new stellar velocity and velocity dispersion 2D maps we measure the radial specific angular momentum profile and compare it with the results from \citet{Foster15}.
  From the stellar metallicity 2D distribution we extract the azimuthally averaged radial metallicity profiles, from which we obtain new measurements for the inner (i.e. $R<1~\rm{R_{e}}$) and outer (i.e. $R\geq1~\rm{R_{e}}$) metallicity gradients and compare them with the results presented in \citet{Pastorello14} and \citet{Forbes15}.

  In this work we assume the following NGC~1023 parameters: effective radius $R_{e}=48~\rm{arcsec}$, axial ratio $b/a=0.37$, position angle $PA=83.3~\rm{degrees}$ and distance $d=11.1~\rm{Mpc}$ \citep{Brodie14}.
  The NGC~1023 photometric decomposition by \citet{Savorgnan15} shows that a faint but spatially extended bar is present in the galaxy inner regions.
  Such a bar extends out to $R_{\rm{bar}}\approx40~\rm{arcsec}$ and has a width of $w_{\rm{bar}}\approx20~\rm{arcsec}$, oriented $\Delta PA = -22~\rm{degrees}$ from the galaxy major axis $PA$.

  The structure of the paper is as follows.
  In Section \ref{maskDesign} we describe the new \textit{SuperSKiMS} mask design and observation method.
  In Section \ref{sec:simulations} we then present how, under typical observing conditions, the data from the \textit{SuperSKiMS} technique would compare with \atlas\ data in the inner regions of NGC~1023.
  In Section \ref{sec:results} we model the galaxy mass density distribution with both the original SKiMS, the \atlas\ and the \textit{SuperSKiMS} mock datasets.
  Section \ref{sec:data} describes the observation and data reduction of two \textit{SuperSKiMS} masks on NGC~1023.
  In Section \ref{dataanalysis} we present the extraction of stellar kinematics and metallicity 2D maps and radial profiles, which we compare with available literature measurements in Section \ref{sec:analysis}.
  In the same Section we compare the results with those from \citet{Foster15} and \citet{Pastorello14}, in order to show the improvement given by the addition of the new data.
  Finally, in Section \ref{sec:conclusions} we present our conclusions.



\section{\textit{SuperSKiMS} mask design}\label{maskDesign}

      As discussed in Section \ref{sec:introduction}, IFU instruments are able to probe the 2D galaxy kinematic and stellar population structures with very high spatial resolution, but they are limited in spatial extent.
      As a consequence, observing a large field-of-view with IFU spectroscopy is time-expensive since it requests a high number of pointings.

      To extend the exploration to larger galactocentric radii in nearby ETGs, one can use longslit or multi-object spectrographs.
      In the first case, the use of simple longslits at multiple position angles allow the exploration out to several effective radii (e.g. \citealt{Statler99, Saglia10}).

      In the second case, one can take advantage of multi-object spectrographs.
        Such instruments generally have a large field-of-view which allows for a wide distribution of slits in 2D.
        The spatial sampling of multislit spectrographs is not comparable with that of IFUs, as the latter naturally sample a 2D field in a contiguous pattern.
      This limitation can be partially overcome by adopting a reliable spatial interpolation technique together with an optimally designed 2D slit distribution.

      Here we present a new multislit mask design, which we call \textit{SuperSKiMS}, that maximizes the azimuthal coverage of the stellar field around the galaxy centre.
      In particular, we test this design on the DEIMOS spectrograph mounted on the Keck II telescope.
      Given a number of DEIMOS masks $N$ used on a given target, the slits in a single mask can be placed within a cone of angle $\alpha=180/N$ degrees.
      In this way, the whole set of masks gives complete azimuthal coverage of the field.

      \placefigI

      In Figure \ref{fig:I} we show the slit distribution of a prototype DEIMOS mask in which the slits are distributed within $45$ degrees cones.
      In this configuration, 4 masks are needed to cover the whole azimuthal range (i.e. $360$ degrees).

      Each SuperSKiMS mask design is adapted to the specific galaxy under observation, given its 2D galaxy surface brightness distribution (i.e. ellipticity and effective radius) and a total exposure time of $2~\rm{hours}$ for each mask.
      The length of each slit is defined from the galaxy surface brightness radial profile in order to obtain a spectrum (after the sky subtraction) with S/N$>8$, which is the threshold to reliably measure the stellar kinematics.
      \citet{Arnold14} showed that a S/N cut off at $\sim$8 generally corresponds to a surface brightness of $\mu_i$ $\sim23~\rm{mag~arcsec^{-2}}$. 
      This cut off allows for DEIMOS spectra kinematic measurements that are commensurate with the estimated kinematics errors.
      Below this S/N ratio threshold, velocity measures may become unreliable.
      In the innermost regions, the minimum slit length is $3$ arcsec.
      In proportion to the galaxy surface brightness profile, larger galactocentric radii slits are longer in order to increase the signal at the expense of the spatial accuracy of the spectra.

      The outermost regions of the mask are reserved for ``sky'' slits or alignment star boxes.
      In the first case, the spectra retrieved from the ``sky'' slits are used to model the sky contribution.
      In addition, between the outer ``sky'' and the inner stellar mask regions, slits targeting globular cluster candidates can be included in the mask design pattern.

      A limitation of DEIMOS is that slits cannot overlap vertically in the mask design. 
      Under this constraint, we need to optimise their spatial distribution in order to maximise the 2D sampling density. 
      This optimal slit distribution in a cone is found via Monte Carlo simulations.
      We simulated $5000$ random slit distributions and, for each of them, we measure the radius of the largest circle (fully within the cone) that can be built in the empty space between the slits. 
      By minimizing this value, we retrieve the slit distribution which has the best spatial sampling.

      The total number of slits in each mask is a free parameter and strongly depends on the surface brightness profile of the galaxy under observation along the mask direction. 
      %



\section{Comparing \textit{SuperSKiMS} mask design with \atlas}\label{sec:simulations}
  As discussed in Section \ref{maskDesign}, issues with the DEIMOS mask alignment software prevented us from properly testing the \textit{SuperSKiMS} mask design.
  As a consequence, the lack of good sampling in the inner regions and the non-optimal coverage of the whole field by our slits prevented us from obtaining 2D kinematic and metallicity maps fully comparable to IFU data.

  \placefigH

  Here, we simulate the sampling of the \textit{SuperSKiMS} central data and how well kriging can recover the underlying field.
  To simulate the potential dataset returned by the \textit{SuperSKiMS} approach in absence of alignment issues, we merge the available standard observations (mostly outside the \atlas\ field-of view) with mocked slits in the central regions (within the \atlas\ field-of-view).
  The mocked slit data are obtained by selecting from the \atlas\ dataset the kinematic and metallicity values at the location of the \textit{SuperSKiMS} slits, assuming a mask with slits within $45$ degree cones observed 4 times (i.e. each time rotated by 45 degrees around the galaxy/mask centre).
  Since most mocked \textit{SuperSKiMS} slits spatially overlap with more than one \atlas\ pixel, to retrieve their values we average the \atlas\ values within the slits' surface, according to the slits' orientation.
  The uncertainties associated with the mocked slits are the sum in quadrature of the uncertainties of all the spatially overlapping \atlas\ measurements.
  We do not include any additional systematic error to account for the use of two different instruments, as our data is consistent with \atlas\ data in the overlapping regions (see Figure \ref{fig:J}).

  In Figure \ref{fig:H} we present a comparison between velocity, velocity dispersion and metallicity 2D maps from the mocked \textit{SuperSKiMS} and \atlas\ data in the innermost region.
  In the \textit{SuperSKiMS} simulation, we build the kriging map from the 43 mocked slits that overlap with the \atlas\ field-of-view.

  In the first column of panels we show the central regions of the \textit{SuperSKiMS} kriged maps and the positions of the mocked slits in the \atlas\ field-of-view.
  In the second column of panels we show the \atlas\ 2D maps at the same spatial scales.
  In order to improve the visualization of such maps, we apply linear interpolation to the \atlas\ datapoints.
  It is worth noticing that, while in the \atlas\ maps a foreground Galactic star is present at $[\Delta \rm{RA}, \Delta \rm{Dec}]\approx[18,10]~\rm{arcsec}$, in the kriged maps such a star is not visible.
  This because no virtual slits have been placed on the star and, therefore, the kriged map does not retrieve such a feature.

  The third column of panels in Figure \ref{fig:H} shows the central region residual maps between the \textit{SuperSKiMS} and the \atlas\ data.
  Similarly to the second column of panels, we apply linear interpolation to the datapoints in order to improve the visualization.
  The residual map values for the velocity are well represented by a Gaussian with a median value of $-2.75~\rm{km~s^{-1}}$ and a standard deviation of $19.49~\rm{km~s^{-1}}$.
  The $rms$ of the residuals is $20.1~\rm{km~s^{-1}}$.

  Similarly, the velocity dispersion residuals show a distribution close to a Gaussian, with a median value of $-1.63~\rm{km~s^{-1}}$ and a standard deviation of $14.17~\rm{km~s^{-1}}$.
  The $rms$ of the velocity dispersion residuals is $14.4~\rm{km~s^{-1}}$.

  Moreover, the stellar metallicity residual distribution is close to a Gaussian with a median value of $-0.01~\rm{dex}$ and a standard deviation of $0.11~\rm{dex}$.
  The $rms$ of the stellar metallicity residuals is $0.11~\rm{dex}$.

  In conclusion, the adoption of the sparsely sampled \textit{SuperSKiMS} technique allows us to retrieve 2D kinematics and metallicity distributions that are highly consistent with a contiguously sampled (and binned) IFU field like \atlas.
  This method can be used with multislit spectrographs other than DEIMOS, which effectively gives them pseudo-IFU capabilities.


\section{Mass modelling with \textit{SuperSKiMS}}
\label{sec:results}

  An important parameter to constrain in galaxy formation models is the fraction of dark matter (DM) in a galaxy \citep{White78b}.
  While in spiral galaxies this quantity can be measured out to large radii from the kinematics of the cold gas, in ETGs such gas is generally not present (although NGC~1023 is an exception, e.g. \citealt{Morganti06}).
  Instead, the fraction of DM in ETGs has been generally measured from the kinematics of the stellar component, which is limited to the bright central regions in most cases, where the stellar mass dominates \citep{Cappellari13a}.

  From massive ETG studies, there is evidence that the total (stellar+DM) mass density profile is consistent with being isothermal, i.e. $\rho_{\rm{tot}}\propto r^{-\gamma_{\rm{tot}}}$ with $\gamma_{\rm{tot}} \approx 2$ (e.g. \citealt{Gerhard01, Thomas11}).
  This is also in agreement with the results from X-ray modelling \citep{Humphrey10} and both strong \citep{Auger10} and weak \citep{Gavazzi07} gravitational lensing.
  However, these studies were biased towards high mass ETGs.

  Recently, \citet{Cappellari15} presented the first homogeneous analysis of the total mass distribution in 14 intermediate mass fast rotator ETGs out to a median radius of $4~\rm{R_{e}}$.
  In this work, the stellar kinematics measured in the galaxy centres from the \atlas\ survey are used together with the outer stellar kinematics obtained by the SLUGGS survey.
  We found that the total mass density in all the galaxies follows a power-law profile, with an average slope $\overline{\gamma}_{\rm{tot}}=2.19\pm0.03$.
  This slope is largely independent of the galaxy mass or central stellar velocity dispersion.

  In \citet{Cappellari15} a source of systematic uncertainties is the adoption of two very different datasets (i.e. \atlas\ and SLUGGS, masks 1 to 5), both in spatial sampling and in uncertainty on the single kinematic measurements.
  In addition, while SLUGGS kinematics are obtained from the Calcium Triplet in the near-infrared, \atlas\ kinematics are measured from optical absorption lines.
  In what follows, we test whether the adoption of a homogeneous dataset obtained from the \textit{SuperSKiMS} technique allows us to retrieve consistent results.
  In particular, we apply the same modelling approach used in \citet{Cappellari15} to two different kinematic datasets for NGC~1023.
  The first dataset (\textit{A+S}) is similar to that used by \citet{Cappellari15} (i.e. combining \atlas\ and SLUGGS kinematics), although it includes the extra slits that have been observed since that work from masks 6 and 7 (see Section \ref{sec:data}).
  The second dataset combines the same SLUGGS kinematic data with the simulated \textit{SuperSKiMS} slits in the inner regions (see Section \ref{sec:simulations}).
  We call this second dataset \textit{S+S}.

  \subsection{Jeans Anisotropic MGE Models}

  With the 2D velocity and velocity dispersion maps it is possible to measure the total mass density profile of NGC~1023, as well as retrieve an estimation of its inclination and orbital anisotropy.
  \citet{Cappellari13a} measured such parameters in the \atlas\ sample for 260 ETGs by constructing Jeans Anisotropic MGE (JAM) dynamical models.
  In their model D, they assumed a spherical DM mass density distribution, following a generalized \citet{Navarro96} (gNFW) radial profile.
  The DM mass density distribution is then:
  \begin{eqnarray}
    \rho_{\rm{DM}}=\rho_{\rm{S}}\left( \frac{r}{r_{\rm{S}}}  \right)^{\gamma_{\rm{DM}}} \left( 0.5+0.5\frac{r}{r_{\rm{S}}}\right)^{-3-\gamma_{\rm{DM}}}
  \end{eqnarray}
  where $\rho_{\rm{S}}$ is the DM mass density at the scale radius $r_{\rm{S}}$ and $\gamma_{\rm{DM}}$ is the slope for $r \ll r_{\rm{S}}$.

    We model the stellar light distribution in 2D using the Multi-Gaussian Expansion (MGE, \citealt{Emsellem94,Cappellari02}) tabulated, for the $i$-band surface brightness, by \citet{Scott09}.
    Unlike \citet{Cappellari15}, we do not clean and symmetrize our initial datasets, although in the \textit{A+S} case a proper weighting of the central points is needed.
    In fact, in this dataset the high number of central points with small uncertainties in the \atlas\ field-of-view would dominate the best-fit model estimation.
    Thus, the values in the outer regions (i.e. where the DM contribution is higher) would be under weighted in the estimation of the DM mass density profile.
    To compensate for this, we artificially increase the uncertainties on both \atlas\ velocity and velocity dispersion values in order to have SLUGGS and \atlas\ datapoints contribute equally to the $\chi^{2}$ value of each evaluated model, as in \citet{Cappellari15}.
    The uncertainties on the \textit{SuperSKiMS} and SLUGGS kinematic values are unchanged.
    The output best-fit models are symmetric in both the \textit{A+S} and \textit{S+S} cases.

    In order to model the stellar kinematics of both the \textit{A+S} and \textit{S+S} datasets, we use the Python version of the JAM code presented in \citet{Cappellari08}.
    We adopt the python module \texttt{emcee} \citep{Foreman-Mackey13} to obtain the posterior probability distribution of the model parameters from a Markov chain Monte Carlo (MCMC) sampling with the Metropolis-Hastings algorithm \citep{Hasting70}.
    The model free parameters are: the galaxy inclination $i$, the inner ($R<\rm{R_{e}}$) and outer ($R>\rm{R_{e}}$) orbital anisotropies $\beta_{\rm{in}}$ and $\beta_{\rm{out}}$, and total mass density profile parameters $\rho_{\rm{S}}$, and $\gamma_{\rm{tot}}$.
    The two different orbital anisotropies are applied to the MGE Gaussians with $\sigma$ (i.e. the width of the Gaussian spatial component) less and greater than $1~\rm{R_{e}}$, respectively \citep{Cappellari15}.
    The scale radius parameter is fixed to $r_{\rm{S}} = 20~\rm{kpc}$ \citep{Cappellari13a}.

    In Figure \ref{fig:K} we show the NGC~1023 $V_{\rm{rms}}$ distribution from both the \textit{A+S} and \textit{S+S} datasets, together with their best fit models.
    In the same Figure we also present the posterior probability distribution from the MCMC sampling for the total mass density slope $\gamma_{\rm{tot}}$.
    In particular, in the \textit{A+S} and \textit{S+S} datasets the best fit model has a total mass density slope of $\gamma_{\rm{tot, A+S}}=-2.03^{+0.35}_{-0.55}$ and $\gamma_{\rm{tot, S+S}}=-2.08^{+0.33}_{-0.44}$, respectively.
    These values are in good agreement with each other and with the $\gamma_{\rm{tot}}$ measured by \citet{Cappellari15} for NGC~1023 (i.e. $\gamma_{\rm{tot,~C15}}=-2.19$).
    They are typical of a galaxy with a nearly isothermal total density distribution.
    Therefore, probing the central regions of NGC~1023 with a \textit{SuperSKiMS} slit pattern (i.e. 4 multi-object spectrograph pointings) allows us to retrieve consistent results with those from traditional IFU spectroscopy.

    In particular, we measure a consistent slope for the total mass density profile.

\placefigK


\section{Observations and Data reduction}\label{sec:data}

            In Table \ref{tab:A} we summarise all the observations from which we obtain the stellar data presented in this paper.

            Masks 1 to 5 have been observed as part of the ongoing SAGES Legacy Unifying Globulars and GalaxieS (SLUGGS) survey\footnote{http://sluggs.swin.edu.au} and the kinematics and metallicity obtained from them have been published by \cite{Arnold14}, \citet{Foster15} and \citet{Pastorello14}.
            The two most recently observed masks (i.e. masks $6$ and $7$) are presented here for the first time.
            These two masks incorporated the \textit{SuperSKiMS} mask design.
            These masks were designed to be aligned along the major and the minor axis of NGC~1023, respectively.
            Unfortunately, during the observation we experienced some issues with the mask alignment software.
            Because of this, we did not obtain a complete azimuthal coverage of the field, although we successfully added 74 spectra to the previously available dataset (particularly in the innermost regions, see Figure \ref{fig:A}).
            Another consequence of the alignment issues is the misalignment 
            between some exposures of the same masks. 
            For this reason, we could not reach the targeted spectral S/N ratio in the outer regions of the galaxy since co-addition of spectra from different exposures was not possible.

            The SLUGGS survey aims to study globular cluster (GC) systems and integrated starlight in a number of nearby ETGs \citep{Brodie14} using specifically designed DEIMOS multislit masks.
            The instrument is set to use a central wavelength of 7800 $\rm{\AA}$ and a grating of 1200 $\rm{line}~\rm{mm^{-1}}$.
            The slits are 1~$\rm{arcsec}$ wide, which yields a resolution of $\Delta \lambda~\sim~1.5~\rm{\AA}$ (i.e. $\Delta V~\sim~50~\rm{km~s^{-1}}$).
            With this setup, we are able to efficiently cover the Calcium Triplet (CaT) region around 8500 $\rm{\AA}$.
            A DEIMOS mask encloses 16.7 $\times$ 5 arcmin$^2$ and typically includes up to $\approx$100 slits.
            Because of this large field-of-view, DEIMOS has been very efficient in obtaining high S/N spectra of GC candidates in nearby galaxies out to large galactocentric radii in few hours of exposure (typically around 2 hours per mask).
            From these slits it is possible to obtain the spectra of both the GC candidates and the background integrated stellar light.
            Throughout this paper we focus on the latter, using DEIMOS to probe the galaxy integrated stellar population kinematics and metallicity.

            \placetabA

            \placefigAsecond

            In Figure \ref{fig:A}, the positions of the slits from which we have been able to extract the galaxy stellar light spectra are shown for all the 7 masks.
            These slits sparsely cover the field out to more than $3~\rm{R_{e}}$.
            However, most of the DEIMOS mask's real estate is not usable to probe the integrated stellar population of NGC~1023.
            This is because the outer regions of the masks are located where the stellar surface brightness is too low for extracting spectra with S/N$>8$ (the measurement of the stellar kinematics in spectra with lower S/N ratios is not reliable).

    \subsection{Data reduction}

        \subsubsection{SKiMS}\label{SKiMS}

            All DEIMOS data presented in this work are reduced with a modified version of the \texttt{SPEC2D} pipeline \citep{Cooper12,Newman13} as described in \citet{Arnold14}.
            In each slit, the pipeline identifies the bright objects (e.g. GC candidates) and it extracts their spectra.
            After this, the spectra from the remaining slit pixels are summed and normalized, to be then used as background light, later subtracted from the bright target spectra.
            This background spectrum contains contributions from the background galaxy integrated stellar light and from the sky.

            It is possible to separate these two contributions with the SKiMS technique, first described in \citet{Norris08, Proctor09} and \citet{Foster09}.
            The first step in this method is the estimation of the sky contribution in the different slits to the total background.
            In order to quantify the sky contribution in each background spectrum, we define a sky index as the ratio of the spectral flux in a sky-dominated band (i.e. $8605.0-8695.5~\rm{\AA}$) to the flux in two nearby sky-free bands ($8526.0-8536.5$ and $8813.0-8822.5~\rm{\AA}$), as described in \citet{Proctor09}.
            This sky index is proportional to the sky contribution over the continuum level (i.e. background stellar light).
            Slits more than several effective radii from the galaxy centre show almost constant sky index values.
            This is an indication that such slits include negligible galaxy light contribution and are hence used as sky spectra.
            In each slit, we model the sky contribution with the penalized maximum likelihood \texttt{pPXF} software \citep{Cappellari04}, using the sky spectra as templates.
            Finally, the best fitting linear combination of these templates in each slit is then subtracted from the original spectrum.



\section{Data Analysis}\label{dataanalysis}

            Using the same software, after sky subtraction, we fit the integrated stellar spectrum with a set of template stars obtained with DEIMOS adopting the same instrumental setup.
            \texttt{pPXF} returns the first four line-of-sight velocity distribution Gauss-Hermite moments and the relative contributions of the templates to the final fitted spectrum.
            In this work we focus on just the first two moments (i.e. mean velocity $v$ and velocity dispersion $\sigma$).
            In addition, from the same spectra we obtain the stellar total metallicity $[Z$/H] (see Section \ref{sec:metallicity}).

            \placefigF

            Finally, we measure the S/N ratio of all spectra as the median of the ratio between the flux and square root of the variance at each wavelength.
            In Figure \ref{fig:F} we present the S/N ratio for all the NGC~1023 stellar spectra in our dataset against the circularised galactocentric radius:
            \begin{eqnarray}
              R_{c} = \sqrt{x^{2}q+y^{2}/q}
            \end{eqnarray}
            where $q$ is the photometric axial ratio and $x, y$ are the coordinates along the galaxy major and minor axes, respectively.
            The slit points are placed on top of the \citet{Noordermeer08} R-band NGC~1023 surface brightness profile and are colour coded according to their S/N.
            In the same plot we show the Maunakea R-band sky surface brightness during a typical SLUGGS observation (i.e. $\mu_{R} = 19.9-20.4~\rm{mag~arcsec^{-2}}$)\footnote{http://www.gemini.edu/sciops/telescopes-and-sites/observing-condition-constraints/optical-sky-background}.
            We obtain stellar spectra from which we retrieve the stellar kinematics out to more than $170~\rm{arcsec}$ (corresponding to almost $3.6~\rm{R_{e}}$), where the galaxy surface brightness is several orders of magnitude fainter than the sky.

        \subsection{2D mapping}

            Since the spectra we obtain from DEIMOS are located at random spatial positions in the sky, in order to obtain reliable 2D maps for the extracted parameters (e.g. velocity, metallicity) we use the kriging technique.
            In particular, we adopt the kriging code included in the package \texttt{fields} \citep{Furrer09} written in the programming language \texttt{R}.
            So far kriging has been used in several SLUGGS papers (e.g. \citealt{Pastorello14,Foster13}), principally because it is able to recover a reliable estimation of the 2D distribution of a variable without any prior assumption.
            An exhaustive description of kriging can be found in \citet{Pastorello14}.
            In this work, we use kriging to obtain 2D maps for the stellar kinematics and the metallicity that will be compared with IFU 2D maps from the \atlas\ survey.
            %

\section{Results}
\label{sec:analysis}

  \subsection{NGC~1023 kinematics and metallicity}\label{SKiMSanalysis}

  In this Section we present the 2D velocity, velocity dispersion and total metallicity maps for NGC~1023 obtained from the whole available dataset (i.e. 7 masks).

  We do not compare our kinematic maps with those presented in \citet{Foster15}, since the latter are based on a largely overlapping dataset, except for the very inner regions. 
  In fact, they do not measure the stellar kinematics for $R<10~\rm{arcsec}$, and their kriging values at such radii are extrapolated from outer region measurements. 
  For this reason, we compare our kinematic measurements with those from the independent literature studies that probed NGC~1023 at these small radii. 

  We compare our results with the available \atlas\ IFU measurements and with the kinematics from the longslit observations by \citet{Debattista02} and \citet{Fabricius12}.
  The \sauron\ kinematics for NGC~1023 were originally presented in \citet{Emsellem04}, but we used the re-analysis of the same data from \citet{Cappellari11a}.
  \citet{Debattista02} measured NGC~1023 stellar velocity and velocity dispersion radial profiles from a longslit placed at $PA=80~\rm{degrees}$ with width $0.7~\rm{arcsec}$.
  Similarly, \citet{Fabricius12} extracted NGC~1023 kinematics from a longslit at $PA=87~\rm{degrees}$ with a width of $1~\rm{arcsec}$.
  In order to compare with these two literature works, we extract virtual slits with similar $PA$s and widths from both the \atlas\ and our kriged 2D maps.

  Finally, we update several results from \citet{Foster15} and \citet{Pastorello14} with the addition of the newly obtained \textit{SuperSKiMS} data.  

    \subsubsection{Stellar velocity}
    In Figure \ref{fig:B} we show the NGC~1023 2D stellar velocity map obtained from our spectra.
    This map has been created using the kriging technique on a sample of $237$ spectra with S/N$\geq8$, and it extends out to almost $3.6~\rm{R_{e}}$.
    Clear signs of stellar rotation are visible out to these large galactocentric radii.

    Most of the slits are randomly distributed in the field and do not cover the most central regions of NGC~1023.
    This is because most of the observed masks were originally targeting GC candidates in NGC~1023 and the high galaxy surface brightness prevents the detection of reliable GC candidates.

    In the second panel of Figure \ref{fig:B} we plot the \citet{Fabricius12} stellar velocity profile, together with the profiles extracted from \atlas and our maps assuming a virtual slit with a width of $1~\rm{arcsec}$ placed at $PA=87~\rm{degrees}$.
    Similarly, in the third panel we compare the radial velocity profile by \citet{Debattista02} with those extracted from \atlas and our maps assuming a virtual slit of width $0.7~\rm{arcsec}$ and $PA=80~\rm{degrees}$.
    The confidence limits for our velocity radial profile are obtained with both Monte Carlo simulations and bootstrapping.
    In the first case, we build 1000 kriging maps from datasets with the same number of data points and the same spatial position, but different associated velocity values.
    These velocities are randomly extracted from a bi-Gaussian distribution, where the positive and the negative $\sigma$ are the positive and negative velocity uncertainties on the measured datapoint, respectively.
    Similarly, in the second case we build 1000 kriging maps from datasets obtained sampling with replacement the original dataset, but keeping the same velocity values and associated uncertainties.
    We obtain a 1D velocity profile from each kriging 2D map.
    From these sets of profiles, we obtain a median value and a distribution of velocity values in each radial bin from which we measure the velocity uncertainty.
    Since this distribution is not a Gaussian, we measure the uncertainties as the $16^{\rm{th}}$ and $84^{\rm{th}}$ percentiles of the distribution.

    Because of the lack of datapoints in the central regions, the radial velocity profile we measure has higher uncertainties in these regions.
    We exclude from our profiles the inner region where we do not have measured data points.

    In both the middle and bottom panels of Figure \ref{fig:B}, the profiles extracted from the kriging maps show larger uncertainties at $R_{\rm{c}}\approx30~\rm{arcsec}$.
    These uncertainties reflect both a lack of measured datapoints near the narrow considered spatial region and the steepness of the spatial gradient of the velocity at those radii.
    This issue does not strongly affect azimuthally averaged profiles, where the higher number of datapoints scattered in the field provides reliable radial measurements (see \citealt{Pastorello14}).

    Overall, we note a good agreement of our kriged 2D map extracted profile with the \atlas\ profile at positive radii, within the uncertainties.
    Similarly, we observe a fair agreement with the two longslit results, within our errorbars at positive radii.
    On the other hand, at negative radii our interpolated values are less consistent with both \atlas\ and longslit profiles.
    These differences may be caused by a lower number of our points in the region of disagreement and, in the \atlas\ case, by the contribution of the NGC~1023 bar.
    In fact, the presence of the bar may be affecting our scattered data points differently than the continuous data point distributions of \atlas.
    For the same reason (i.e. fewer data points), the kriging profile could suffer by possible interpolation errors.
    This would explain why, despite the profiles' slight disagreement, our and \atlas\ individual velocity data points at the same locations are in very good agreement (see Section \ref{ComparisonSKiMS_Atlas}).

    \placefigB

    \subsubsection{Stellar velocity dispersion}
    In Figure \ref{fig:C} we show the NGC~1023 velocity dispersion 2D map obtained from our spectra.
    Similarly to the map in Figure \ref{fig:B}, to create this map we have used the kriging technique with all the $237$ available data points with S/N$\geq8$.
    We detect a gradient of the velocity dispersion, with central regions showing higher values of $\sigma$ than the outer regions.
    However, the lack of slits near the very centre of the galaxy prevents the observation of a clear high velocity dispersion peak.

    In the second panel of Figure \ref{fig:C} we compare the velocity dispersion profile from \citet{Fabricius12} with the profiles extracted from virtual slits on \atlas\ and our kriging maps.
    These virtual slits have the same orientation and width as the actual slit in comparison.
    Similarly, in the third panel we compare the velocity dispersion profile of \citet{Debattista02} with those extracted from \atlas\ and our kriged 2D map virtual slits with the same $PA$ and width.
    We obtain the velocity dispersion profile's uncertainties in a similar fashion to the velocity profile case presented above.

    In general, we observe a remarkable consistency between our virtual slit radial profile and that from \atlas\ in the overlapping radial regions, although these regions are small.
    However, in the outermost regions (i.e. outside $\approx30~\rm{arcsec}$ in both directions), the longslit profiles show higher stellar velocity dispersions.
    Such disagreements are similar in both the negative and positive sides of the profiles.
    Therefore, a possible cause is the increasingly larger spatial binning that is applied to the outer regions of the longslits in order to preserve the S/N ratio of the extracted spectra.
    As a consequence, such bins include information from stars with different radial velocities.
    This may increase the measured velocity dispersion with respect to the real stellar velocity dispersion along the line of sight, although we note that the velocity profile at such radii is quite flat.
    We suspect that a second possible contribution to the velocity dispersion offset might be linked with the radii at which the literature binned values are assigned.
    If such radii are not weighted according to the galaxy surface brightness profile within the bin, they would overestimate the galactocentric radii where the stellar velocity dispersion has been measured.
    A third possibility is that the offset is linked to the different wavelength regions from which the kinematics are measured.
    However, since our velocity dispersion measurements (i.e. from the CaT lines in the near-infrared) are very consistent with those from \atlas\ (i.e. from absorption lines in the optical), we consider this possibility as unlikely.

    \placefigC

    \subsubsection{Comparison between \atlas\ and SKiMS kinematics}\label{ComparisonSKiMS_Atlas}
    The comparison between the velocity dispersion measurements by \atlas\ and those obtained from the SKiMS technique from masks 1 to 5 at the same spatial locations  shows a systematic offset in most of the galaxies in common between the two datasets \citep{Arnold14, Foster15}.
    In particular, \citet{Foster15} found that the difference between \atlas\ and SKiMS velocity dispersions is on average $\Delta \sigma \approx 20~\rm{km~s^{-1}}$, while the other velocity moments (i.e. $v$, $h_{3}$ and $h_{4}$) show a good general agreement.

    \placefigJ

    In Figure \ref{fig:J} we show the comparison between the velocity ($v$), velocity dispersion ($\sigma$), $h_{3}$ and $h_{4}$ measurements of our updated dataset (i.e. masks 1 to 7) and those from \atlas.
    In particular, we compare our datapoints with those from \atlas\ at the same spatial locations within $1~\rm{arcsec}$.
    Most of \atlas\ values at their largest radii are measured from spectra that have been spatially binned in order to increase their S/N ratio.
    We use the centre coordinates of these bins to evaluate their distance from our points.
    As a consequence, our points in outer \atlas\ regions are generally not matched with their strongly-binned \atlas\ counterparts (since they are generally not within $1~\rm{arcsec}$).
    In total, we have 18 slits within the \atlas\ field-of-view, of which 8 are within $1~\rm{arcsec}$ of an \atlas\ kinematic measurement.

    To the velocity and velocity dispersion uncertainties obtained with SKiMS we add in quadrature $5~\rm{km~s^{-1}}$ and $8~\rm{km~s^{-1}}$, respectively, to take into account systematics (see \citealt{Foster11} for a justification).

    As in \citet{Foster15}, we find a good match between our velocity measurements and those from \atlas.
    Comparing the two datasets, we find an average offset of $\Delta v = 9.64 \pm0.94~\rm{km~s^{-1}}$, well below the velocity resolution of both instruments.

    However, \citet{Foster15} found an average offset between SKiMS and \atlas\ velocity dispersions in NGC~1023 of $\Delta \sigma \approx 20~\rm{km~s^{-1}}$.
    In the \textit{top~right} panel of Figure \ref{fig:J} we plot our velocity dispersion measurements against the spatially overlapping measurements in \atlas.
    We do not find a strong offset with the \atlas\ velocity dispersion measurements, although the points present a large scatter.
    In particular, we measure an average difference of $\Delta \sigma = 0.88 \pm 1.21~\rm{km~s^{-1}}$.
    This better agreement is probably due to the larger dataset we use, that probes more into the central regions of NGC~1023 (i.e. where the \atlas\ datapoints are less binned), although we cannot exclude the presence of other systematics.
    At the same time, our comparison tends to exclude the points in the strongly binned regions near the edges of the \atlas\ field-of-view.

    The \textit{bottom~left} and \textit{bottom~right} panels of Figure \ref{fig:J} show the comparison of \atlas\ and SKiMS measurements for the third (i.e. $h_{3}$) and fourth (i.e. $h_{4}$) stellar velocity moments, respectively.
    In both cases there is a fair agreement between the two datasets with the exception of few outliers with large uncertainties.
    The average offset between our $h_{3}$ measurements and those by \atlas\ is $\Delta h_{3} = -0.005 \pm 0.058$, consistent with being zero.
    Similarly, the average offset between the two datasets for the $h_{4}$ values is $\Delta h_{4} = -0.03 \pm 0.06$.

    In all the comparisons, the scatter appears to be mostly due to the few slits of ours that have been matched with the \atlas\ values from large bins (i.e. near the edges of the \atlas\ field-of-view).

    We speculate that the velocity dispersion offset measured by \citet{Foster15} for NGC~1023 might be caused by the lack of datapoints in the innermost regions of the galaxy.
    However, in other galaxies in common between SLUGGS and \atlas\ samples for which a high number of overlapping points already exist, such a possible explanation might not be valid.

    \subsubsection{Local specific angular momentum profile}

    From the velocity and velocity dispersion measurements, \citet{Foster15} extracted the local specific angular momentum profile $\lambda(R)$ for a number of galaxies, including NGC~1023.
    In this work we present new data for NGC~1023, which extend more towards the centre of the galaxy.
    From our velocity and velocity dispersion profiles, we measure:
    \begin{eqnarray}
    \lambda(R) = \frac{|V(R)|}{\sqrt{V(R)^{2}+\sigma(R)^{2}}}
    \end{eqnarray}
    where $V(R)$ and $\sigma(R)$ are the velocity and the velocity dispersion at the circularised radius $R$.
    Similarly, we measure the same quantity from the kinematic radial profiles extracted from the kriging maps in \citet{Foster15} (but assuming NGC~1023 photometric parameters presented here in Section \ref{sec:introduction}).

    \placefigL

    In Figure \ref{fig:L} we show both the $\lambda(R)$ radial profiles from the kinematic maps of this work and of \citet{Foster15}.
    From the plot it is possible to appreciate how the addition of new data slightly decreases the uncertainties and allows the exploration of the innermost regions of NGC~1023, while being fully consistent in the outer regions.

    \subsubsection{Metallicity}\label{sec:metallicity}

    From our stellar spectra it is also possible to obtain the stellar total metallicity $[Z$/H].
    This is because the CaT equivalent width strongly correlates with the metallicity of the stellar population and is almost independent of the stellar age if the stellar population is $>2~\rm{Gyr}$ (see \citealt{Usher12} and references therein).
    Similarly to \citet{Pastorello14}, we extract a relation between the CaT equivalent width and the stellar metallicity from the MILES stellar population models \citep{Vazdekis10} assuming a \citet{Salpeter55} initial mass function (IMF).
    Since the CaT equivalent width depends on the IMF slope, we apply the same correction to the CaT-derived metallicity values as in \citet{Pastorello14}.

    In Figure \ref{fig:D} we show the NGC~1023 stellar metallicity map and radial profile.
    The kriging map shows a centrally peaked 2D metallicity distribution, with metallicity isocurves roughly following the galaxy isophotes.

    To retrieve reliable metallicity measurements from the CaT we need a S/N$\geq30$.
    Thus our metallicity map extends to smaller radii than in the stellar kinematics case, i.e. about $2.5~\rm{R_{e}}$ from the galaxy centre.
    The extracted metallicity radial profile is slightly less extended.
    This is because a sufficiently high number of map points are needed in each radial bin in order to obtain statistically reliable uncertainties.

    \placefigD

    In the bottom panel of Figure \ref{fig:D} we compare this azimuthally averaged radial metallicity profiles obtained from our SKiMS (i.e. masks from 1 to 7) kriged 2D map to the \atlas\ dataset from \citet{Kuntschner10}.
    In both cases, we bin the data points in annulii of $1~\rm{arcsec}$ that follow the galaxy isophotal shape and orientation.
    In particular, we adopt radially constant $PA$ and axial ratio (see Section \ref{sec:introduction}).
    The uncertainty on the azimuthally averaged kriging metallicity profile is obtained with the same approach used in the kinematic profiles.

    The radial metallicity profiles shows a remarkable overall agreement, with our profile extending out to almost twice the galactocentric radius probed by \atlas.
    We note that, while our values are obtained from the CaT lines in the near-infrared, \atlas\ metallicities are obtained from optical absorption lines using the Lick system \citep{Kuntschner10}.
    In the central-most regions, our lack of datapoints prevents the kriging map from correctly reproducing the steep metallicity profile that is visible in the \atlas\ profile.

    \citet{Pastorello14} used the radial metallicity profiles extracted from the kriging maps to measure the inner (i.e. $R<1~\rm{R_{e}}$) and outer (i.e. $R>1~\rm{R_{e}}$) metallicity gradients of NGC~1023 in a logarithmic radial scale.
    An updated version of those measurements is presented in the appendix B of \citet{Forbes16}.
    Adopting the same approach but with the updated dataset, we extract new inner and outer stellar metallicity gradients.
    Both \citet{Forbes16} and our metallicity gradients are presented in Table \ref{tab:C}.

    Such new gradients are steeper than those presented in \citet{Forbes16}, although still consistent within the uncertainties. 
    The larger uncertainties for the inner gradient are due to a spatially larger inner region from which such a gradient has been measured. 
    The availability of measured \textit{SuperSKiMS} datapoints in the inner regions allows us to probe the metallicity to smaller radii, where the metallicity profiles becomes steeper and the gradient uncertainty larger. 

    \placetabC


\section{Conclusions}
\label{sec:conclusions}

We present for the first time a new multislit mask design, called \textit{SuperSKiMS}, that maximizes the azimuthal and radial coverage of a galaxy's 2D stellar field.
  This slit mask design can be optimised to the best azimuthal coverage given the expected number of mask pointings.
  In addition, the slit lengths are designed in order to provide a high S/N ratio at most radii.
  The outermost regions of the mask are used to obtain pure-sky spectra necessary for sky subtraction and to target globular cluster candidates or background galaxies.

  We tested the technique by simulating the exposure of four \textit{SuperSKiMS} masks in the central regions of NGC~1023.
  For this galaxy, \atlas\ 2D kinematic and metallicity maps are available.
  We extracted the velocity, velocity dispersion and metallicity measurements from these \atlas\ maps, at the spatial locations where the \textit{SuperSKiMS} central slits would be.
  We obtained 2D maps which we compared with those from \atlas\ in the inner regions.
  We found that the observation of 4 \textit{SuperSKiMS} DEIMOS masks would allow us to recover the stellar kinematic and metallicity 2D distribution observed by the \atlas\ work with high accuracy.

  Moreover, we compared a partially simulated SLUGGS+\textit{SuperSKIMS} dataset with the composite SLUGGS+\atlas\ dataset used by \citet{Cappellari15} to model the total mass density profile of NGC~1023.
  In particular, from both datasets we retrieved a nearly isothermal total mass density distribution in the probed radial range, consistent with that published by \citet{Cappellari15} for this galaxy.

  As a following step, we tested the \textit{SuperSKiMS} technique using the large field-of-view DEIMOS multislit spectrograph on the nearby galaxy NGC~1023.
  Unfortunately, issues with the Keck alignment software prevented us from fully exploiting the capabilities of the \textit{SuperSKiMS} mask design.
  In any case, we observed two masks with the \textit{SuperSKiMS} design and added the new data to the pre-existing SLUGGS dataset.
  With such data we then produced 2D stellar velocity and velocity dispersion maps.
  Such maps allowed the extraction of virtual slit radial profiles, which we compared with longslit literature results.
  Despite random sampling of the field, we found a good consistency with the literature velocity profiles.
  In addition, comparing our kinematic radial profiles with those extracted from \atlas\ 2D maps, we found fair consistency in the overlapping regions.

  Since many of the new datapoints lie in the \atlas\ field-of-view, we also compared the differences in kinematics at the same spatial locations with \atlas\ data.
  We found a good agreement between the velocity and a fair agreement between the velocity dispersion values (in particular, we did not observe the velocity dispersion offset noted in \citealt{Arnold14} and \citealt{Foster15}).
  These results suggest that the previously observed offset in NGC~1023 might be driven by the small number of datapoints overlapping with the \atlas\ field-of-view, most of which also lie at the edges of this region, where the \atlas\ values have been obtained from the binning of low S/N data.
  However, other galaxies in \citet{Arnold14} and \citet{Foster15} show similar offsets even in the presence of a large number of datapoints overlapping with the \atlas\ field-of-view.

  From these kinematic measurements we also extracted the local specific angular momentum radial profile of NGC~1023, extending the results by \citet{Foster15} with our updated dataset.

  In addition to the kinematics, it is possible to reliably measure the equivalent width of the CaT absorption lines and hence produce 2D metallicity maps.
  From these maps we extracted the azimuthally averaged radial metallicity profile out to $R\approx~2\rm{R_{e}}$.
  This profile is consistent with the stellar metallicity measurements by \atlas\ in the overlapping radial regions, while extending to almost twice their probed galactocentric radii.
  In addition, we extracted the inner and outer stellar metallicity gradients for NGC~1023, updating the results already presented in \citet{Forbes16} with our new dataset.

  In future studies, the adoption of the \textit{SuperSKiMS} mask design to observe nearby galaxies with multislit spectrographs will allow one to probe the stellar component homogeneously from the innermost to the outermost regions.
  From this it will be possible to efficiently retrieve reliable kinematics and stellar population 2D maps comparable to those from integral field unit spectrograph studies, although reaching much larger radii.

\section*{Acknowledgements}
	Twe authors want to thank the (anonymous) referee for the useful comments
	and suggestions which helped to improve this paper.
	The authors wish also to express their gratitude to Joachim Janz for his help and feedback.
	Some of the data presented herein were obtained at the W. M. Keck Observatory, operated as a scientific partnership among the California Institute of Technology, the University of California and the National Aeronautics and Space Administration, and made possible by the generous financial support of the W. M. Keck Foundation.
	The authors wish to recognize and acknowledge the very significant cultural role and reverence that the summit of Mauna Kea has always had within the indigenous Hawaiian community.
	The analysis pipeline used to reduce the DEIMOS data was developed at UC Berkeley with support from NSF grant AST-0071048.
	DF thanks the ARC for support via DP130100388.
  AP thanks the AAO for financial support via the Honours/Masters Scholarship.
  MC acknowledges support from a Royal Society University Research Fellowship.
	This work was also supported by NSF grant AST-1211995.

\bibliographystyle{apj}
\bibliography{bibliography}{}

\begin{thebibliography}{59}
\expandafter\ifx\csname natexlab\endcsname\relax\def\natexlab#1{#1}\fi

\bibitem[{{Arnold} {et~al.}(2014){Arnold}, {Romanowsky}, {Brodie}, {Forbes},
  {Strader}, {Spitler}, {Foster}, {Blom}, {Kartha}, {Pastorello}, {Pota},
  {Usher}, \& {Woodley}}]{Arnold14}
{Arnold}, J.~A., {Romanowsky}, A.~J., {Brodie}, J.~P., {Forbes}, D.~A.,
  {Strader}, J., {Spitler}, L.~R., {Foster}, C., {Blom}, C., {Kartha}, S.~S.,
  {Pastorello}, N., {Pota}, V., {Usher}, C., \& {Woodley}, K.~A. 2014, \apj,
  791, 80

\bibitem[{{Arnold} {et~al.}(1994){Arnold}, {de Zeeuw}, \& {Hunter}}]{Arnold94}
{Arnold}, R., {de Zeeuw}, P.~T., \& {Hunter}, C. 1994, \mnras, 271, 924

\bibitem[{{Auger} {et~al.}(2010){Auger}, {Treu}, {Gavazzi}, {Bolton},
  {Koopmans}, \& {Marshall}}]{Auger10}
{Auger}, M.~W., {Treu}, T., {Gavazzi}, R., {Bolton}, A.~S., {Koopmans},
  L.~V.~E., \& {Marshall}, P.~J. 2010, \apjl, 721, L163

\bibitem[{{Bacon} {et~al.}(2001){Bacon}, {Copin}, {Monnet}, {Miller},
  {Allington-Smith}, {Bureau}, {Carollo}, {Davies}, {Emsellem}, {Kuntschner},
  {Peletier}, {Verolme}, \& {de Zeeuw}}]{Bacon01}
{Bacon}, R., {Copin}, Y., {Monnet}, G., {Miller}, B.~W., {Allington-Smith},
  J.~R., {Bureau}, M., {Carollo}, C.~M., {Davies}, R.~L., {Emsellem}, E.,
  {Kuntschner}, H., {Peletier}, R.~F., {Verolme}, E.~K., \& {de Zeeuw}, P.~T.
  2001, \mnras, 326, 23

\bibitem[{{Brodie} {et~al.}(2014){Brodie}, {Romanowsky}, {Strader}, {Forbes},
  {Foster}, {Jennings}, {Pastorello}, {Pota}, {Usher}, {Blom}, {Kader},
  {Roediger}, {Spitler}, {Villaume}, {Arnold}, {Kartha}, \&
  {Woodley}}]{Brodie14}
{Brodie}, J.~P., {Romanowsky}, A.~J., {Strader}, J., {Forbes}, D.~A., {Foster},
  C., {Jennings}, Z.~G., {Pastorello}, N., {Pota}, V., {Usher}, C., {Blom}, C.,
  {Kader}, J., {Roediger}, J.~C., {Spitler}, L.~R., {Villaume}, A., {Arnold},
  J.~A., {Kartha}, S.~S., \& {Woodley}, K.~A. 2014, \apj, 796, 52

\bibitem[{{Bryant} {et~al.}(2015){Bryant}, {Owers}, {Robotham}, {Croom},
  {Driver}, {Drinkwater}, {Lorente}, {Cortese}, {Scott}, {Colless}, {Schaefer},
  {Taylor}, {Konstantopoulos}, {Allen}, {Baldry}, {Barnes}, {Bauer},
  {Bland-Hawthorn}, {Bloom}, {Brooks}, {Brough}, {Cecil}, {Couch}, {Croton},
  {Davies}, {Ellis}, {Fogarty}, {Foster}, {Glazebrook}, {Goodwin}, {Green},
  {Gunawardhana}, {Hampton}, {Ho}, {Hopkins}, {Kewley}, {Lawrence},
  {Leon-Saval}, {Leslie}, {McElroy}, {Lewis}, {Liske}, {L{\'o}pez-S{\'a}nchez},
  {Mahajan}, {Medling}, {Metcalfe}, {Meyer}, {Mould}, {Obreschkow}, {O'Toole},
  {Pracy}, {Richards}, {Shanks}, {Sharp}, {Sweet}, {Thomas}, {Tonini}, \&
  {Walcher}}]{Bryant15}
{Bryant}, J.~J., {Owers}, M.~S., {Robotham}, A.~S.~G., {Croom}, S.~M.,
  {Driver}, S.~P., {Drinkwater}, M.~J., {Lorente}, N.~P.~F., {Cortese}, L.,
  {Scott}, N., {Colless}, M., {Schaefer}, A., {Taylor}, E.~N.,
  {Konstantopoulos}, I.~S., {Allen}, J.~T., {Baldry}, I., {Barnes}, L.,
  {Bauer}, A.~E., {Bland-Hawthorn}, J., {Bloom}, J.~V., {Brooks}, A.~M.,
  {Brough}, S., {Cecil}, G., {Couch}, W., {Croton}, D., {Davies}, R., {Ellis},
  S., {Fogarty}, L.~M.~R., {Foster}, C., {Glazebrook}, K., {Goodwin}, M.,
  {Green}, A., {Gunawardhana}, M.~L., {Hampton}, E., {Ho}, I.-T., {Hopkins},
  A.~M., {Kewley}, L., {Lawrence}, J.~S., {Leon-Saval}, S.~G., {Leslie}, S.,
  {McElroy}, R., {Lewis}, G., {Liske}, J., {L{\'o}pez-S{\'a}nchez}, {\'A}.~R.,
  {Mahajan}, S., {Medling}, A.~M., {Metcalfe}, N., {Meyer}, M., {Mould}, J.,
  {Obreschkow}, D., {O'Toole}, S., {Pracy}, M., {Richards}, S.~N., {Shanks},
  T., {Sharp}, R., {Sweet}, S.~M., {Thomas}, A.~D., {Tonini}, C., \& {Walcher},
  C.~J. 2015, \mnras, 447, 2857

\bibitem[{{Bundy} {et~al.}(2015){Bundy}, {Bershady}, {Law}, {Yan}, {Drory},
  {MacDonald}, {Wake}, {Cherinka}, {S{\'a}nchez-Gallego}, {Weijmans}, {Thomas},
  {Tremonti}, {Masters}, {Coccato}, {Diamond-Stanic}, {Arag{\'o}n-Salamanca},
  {Avila-Reese}, {Badenes}, {Falc{\'o}n-Barroso}, {Belfiore}, {Bizyaev},
  {Blanc}, {Bland-Hawthorn}, {Blanton}, {Brownstein}, {Byler}, {Cappellari},
  {Conroy}, {Dutton}, {Emsellem}, {Etherington}, {Frinchaboy}, {Fu}, {Gunn},
  {Harding}, {Johnston}, {Kauffmann}, {Kinemuchi}, {Klaene}, {Knapen},
  {Leauthaud}, {Li}, {Lin}, {Maiolino}, {Malanushenko}, {Malanushenko}, {Mao},
  {Maraston}, {McDermid}, {Merrifield}, {Nichol}, {Oravetz}, {Pan}, {Parejko},
  {Sanchez}, {Schlegel}, {Simmons}, {Steele}, {Steinmetz}, {Thanjavur},
  {Thompson}, {Tinker}, {van den Bosch}, {Westfall}, {Wilkinson}, {Wright},
  {Xiao}, \& {Zhang}}]{Bundy15}
{Bundy}, K., {Bershady}, M.~A., {Law}, D.~R., {Yan}, R., {Drory}, N.,
  {MacDonald}, N., {Wake}, D.~A., {Cherinka}, B., {S{\'a}nchez-Gallego}, J.~R.,
  {Weijmans}, A.-M., {Thomas}, D., {Tremonti}, C., {Masters}, K., {Coccato},
  L., {Diamond-Stanic}, A.~M., {Arag{\'o}n-Salamanca}, A., {Avila-Reese}, V.,
  {Badenes}, C., {Falc{\'o}n-Barroso}, J., {Belfiore}, F., {Bizyaev}, D.,
  {Blanc}, G.~A., {Bland-Hawthorn}, J., {Blanton}, M.~R., {Brownstein}, J.~R.,
  {Byler}, N., {Cappellari}, M., {Conroy}, C., {Dutton}, A.~A., {Emsellem}, E.,
  {Etherington}, J., {Frinchaboy}, P.~M., {Fu}, H., {Gunn}, J.~E., {Harding},
  P., {Johnston}, E.~J., {Kauffmann}, G., {Kinemuchi}, K., {Klaene}, M.~A.,
  {Knapen}, J.~H., {Leauthaud}, A., {Li}, C., {Lin}, L., {Maiolino}, R.,
  {Malanushenko}, V., {Malanushenko}, E., {Mao}, S., {Maraston}, C.,
  {McDermid}, R.~M., {Merrifield}, M.~R., {Nichol}, R.~C., {Oravetz}, D.,
  {Pan}, K., {Parejko}, J.~K., {Sanchez}, S.~F., {Schlegel}, D., {Simmons}, A.,
  {Steele}, O., {Steinmetz}, M., {Thanjavur}, K., {Thompson}, B.~A., {Tinker},
  J.~L., {van den Bosch}, R.~C.~E., {Westfall}, K.~B., {Wilkinson}, D.,
  {Wright}, S., {Xiao}, T., \& {Zhang}, K. 2015, \apj, 798, 7

\bibitem[{{Cappellari}(2002)}]{Cappellari02}
{Cappellari}, M. 2002, \mnras, 333, 400

\bibitem[{{Cappellari}(2008)}]{Cappellari08}
---. 2008, \mnras, 390, 71

\bibitem[{{Cappellari} \& {Emsellem}(2004)}]{Cappellari04}
{Cappellari}, M. \& {Emsellem}, E. 2004, \pasp, 116, 138

\bibitem[{{Cappellari} {et~al.}(2011){Cappellari}, {Emsellem}, {Krajnovi{\'c}},
  {McDermid}, {Scott}, {Verdoes Kleijn}, {Young}, {Alatalo}, {Bacon}, {Blitz},
  {Bois}, {Bournaud}, {Bureau}, {Davies}, {Davis}, {de Zeeuw}, {Duc},
  {Khochfar}, {Kuntschner}, {Lablanche}, {Morganti}, {Naab}, {Oosterloo},
  {Sarzi}, {Serra}, \& {Weijmans}}]{Cappellari11a}
{Cappellari}, M., {Emsellem}, E., {Krajnovi{\'c}}, D., {McDermid}, R.~M.,
  {Scott}, N., {Verdoes Kleijn}, G.~A., {Young}, L.~M., {Alatalo}, K., {Bacon},
  R., {Blitz}, L., {Bois}, M., {Bournaud}, F., {Bureau}, M., {Davies}, R.~L.,
  {Davis}, T.~A., {de Zeeuw}, P.~T., {Duc}, P.-A., {Khochfar}, S.,
  {Kuntschner}, H., {Lablanche}, P.-Y., {Morganti}, R., {Naab}, T.,
  {Oosterloo}, T., {Sarzi}, M., {Serra}, P., \& {Weijmans}, A.-M. 2011, \mnras,
  413, 813

\bibitem[{{Cappellari} {et~al.}(2013{\natexlab{a}}){Cappellari}, {McDermid},
  {Alatalo}, {Blitz}, {Bois}, {Bournaud}, {Bureau}, {Crocker}, {Davies},
  {Davis}, {de Zeeuw}, {Duc}, {Emsellem}, {Khochfar}, {Krajnovi{\'c}},
  {Kuntschner}, {Morganti}, {Naab}, {Oosterloo}, {Sarzi}, {Scott}, {Serra},
  {Weijmans}, \& {Young}}]{Cappellari13b}
{Cappellari}, M., {McDermid}, R.~M., {Alatalo}, K., {Blitz}, L., {Bois}, M.,
  {Bournaud}, F., {Bureau}, M., {Crocker}, A.~F., {Davies}, R.~L., {Davis},
  T.~A., {de Zeeuw}, P.~T., {Duc}, P.-A., {Emsellem}, E., {Khochfar}, S.,
  {Krajnovi{\'c}}, D., {Kuntschner}, H., {Morganti}, R., {Naab}, T.,
  {Oosterloo}, T., {Sarzi}, M., {Scott}, N., {Serra}, P., {Weijmans}, A.-M., \&
  {Young}, L.~M. 2013{\natexlab{a}}, \mnras, 432, 1862

\bibitem[{{Cappellari} {et~al.}(2015){Cappellari}, {Romanowsky}, {Brodie},
  {Forbes}, {Strader}, {Foster}, {Kartha}, {Pastorello}, {Pota}, {Spitler},
  {Usher}, \& {Arnold}}]{Cappellari15}
{Cappellari}, M., {Romanowsky}, A.~J., {Brodie}, J.~P., {Forbes}, D.~A.,
  {Strader}, J., {Foster}, C., {Kartha}, S.~S., {Pastorello}, N., {Pota}, V.,
  {Spitler}, L.~R., {Usher}, C., \& {Arnold}, J.~A. 2015, \apjl, 804, L21

\bibitem[{{Cappellari} {et~al.}(2013{\natexlab{b}}){Cappellari}, {Scott},
  {Alatalo}, {Blitz}, {Bois}, {Bournaud}, {Bureau}, {Crocker}, {Davies},
  {Davis}, {de Zeeuw}, {Duc}, {Emsellem}, {Khochfar}, {Krajnovi{\'c}},
  {Kuntschner}, {McDermid}, {Morganti}, {Naab}, {Oosterloo}, {Sarzi}, {Serra},
  {Weijmans}, \& {Young}}]{Cappellari13a}
{Cappellari}, M., {Scott}, N., {Alatalo}, K., {Blitz}, L., {Bois}, M.,
  {Bournaud}, F., {Bureau}, M., {Crocker}, A.~F., {Davies}, R.~L., {Davis},
  T.~A., {de Zeeuw}, P.~T., {Duc}, P.-A., {Emsellem}, E., {Khochfar}, S.,
  {Krajnovi{\'c}}, D., {Kuntschner}, H., {McDermid}, R.~M., {Morganti}, R.,
  {Naab}, T., {Oosterloo}, T., {Sarzi}, M., {Serra}, P., {Weijmans}, A.-M., \&
  {Young}, L.~M. 2013{\natexlab{b}}, \mnras, 432, 1709

\bibitem[{{Cooper} {et~al.}(2012){Cooper}, {Newman}, {Davis}, {Finkbeiner}, \&
  {Gerke}}]{Cooper12}
{Cooper}, M.~C., {Newman}, J.~A., {Davis}, M., {Finkbeiner}, D.~P., \& {Gerke},
  B.~F. 2012, Astrophysics Source Code Library, 3003

\bibitem[{{Cretton} {et~al.}(2000){Cretton}, {Rix}, \& {de Zeeuw}}]{Cretton00}
{Cretton}, N., {Rix}, H.-W., \& {de Zeeuw}, P.~T. 2000, \apj, 536, 319

\bibitem[{{Davies} \& {Birkinshaw}(1988)}]{Davies88}
{Davies}, R.~L. \& {Birkinshaw}, M. 1988, \apjs, 68, 409

\bibitem[{{Debattista} {et~al.}(2002){Debattista}, {Corsini}, \&
  {Aguerri}}]{Debattista02}
{Debattista}, V.~P., {Corsini}, E.~M., \& {Aguerri}, J.~A.~L. 2002, \mnras,
  332, 65

\bibitem[{{Emsellem} {et~al.}(2004){Emsellem}, {Cappellari}, {Peletier},
  {McDermid}, {Bacon}, {Bureau}, {Copin}, {Davies}, {Krajnovi{\'c}},
  {Kuntschner}, {Miller}, \& {de Zeeuw}}]{Emsellem04}
{Emsellem}, E., {Cappellari}, M., {Peletier}, R.~F., {McDermid}, R.~M.,
  {Bacon}, R., {Bureau}, M., {Copin}, Y., {Davies}, R.~L., {Krajnovi{\'c}}, D.,
  {Kuntschner}, H., {Miller}, B.~W., \& {de Zeeuw}, P.~T. 2004, \mnras, 352,
  721

\bibitem[{{Emsellem} {et~al.}(1994){Emsellem}, {Monnet}, \&
  {Bacon}}]{Emsellem94}
{Emsellem}, E., {Monnet}, G., \& {Bacon}, R. 1994, \aap, 285, 723

\bibitem[{{Fabricius} {et~al.}(2012){Fabricius}, {Saglia}, {Fisher}, {Drory},
  {Bender}, \& {Hopp}}]{Fabricius12}
{Fabricius}, M.~H., {Saglia}, R.~P., {Fisher}, D.~B., {Drory}, N., {Bender},
  R., \& {Hopp}, U. 2012, \apj, 754, 67

\bibitem[{{Forbes} {et~al.}(2015){Forbes}, {Pastorello}, {Romanowsky}, {Usher},
  {Brodie}, \& {Strader}}]{Forbes15}
{Forbes}, D.~A., {Pastorello}, N., {Romanowsky}, A.~J., {Usher}, C., {Brodie},
  J.~P., \& {Strader}, J. 2015, \mnras, 452, 1045

\bibitem[{{Forbes} {et~al.}(2016){Forbes}, {Romanowsky}, {Pastorello},
  {Foster}, {Brodie}, {Strader}, {Pota}, \& {Usher}}]{Forbes16}
{Forbes}, D.~A., {Romanowsky}, A.~J., {Pastorello}, N., {Foster}, C., {Brodie},
  J.~P., {Strader}, J., {Pota}, V., \& {Usher}, C. 2016, \mnras

\bibitem[{{Foreman-Mackey} {et~al.}(2013){Foreman-Mackey}, {Hogg}, {Lang}, \&
  {Goodman}}]{Foreman-Mackey13}
{Foreman-Mackey}, D., {Hogg}, D.~W., {Lang}, D., \& {Goodman}, J. 2013, \pasp,
  125, 306

\bibitem[{{Foster} {et~al.}(2013){Foster}, {Arnold}, {Forbes}, {Pastorello},
  {Romanowsky}, {Spitler}, {Strader}, \& {Brodie}}]{Foster13}
{Foster}, C., {Arnold}, J.~A., {Forbes}, D.~A., {Pastorello}, N., {Romanowsky},
  A.~J., {Spitler}, L.~R., {Strader}, J., \& {Brodie}, J.~P. 2013, \mnras, 435,
  3587

\bibitem[{{Foster} {et~al.}(2015){Foster}, {Pastorello}, {Roediger}, {Arnold},
  {Brodie}, {Forbes}, {Romanowsky}, \& {Spitler}}]{Foster15}
{Foster}, C., {Pastorello}, N., {Roediger}, J., {Arnold}, J.~A., {Brodie},
  J.~P., {Forbes}, D.~A., {Romanowsky}, A.~J., \& {Spitler}, L.~R. 2015, \mnras

\bibitem[{{Foster} {et~al.}(2009){Foster}, {Proctor}, {Forbes}, {Spolaor},
  {Hopkins}, \& {Brodie}}]{Foster09}
{Foster}, C., {Proctor}, R.~N., {Forbes}, D.~A., {Spolaor}, M., {Hopkins},
  P.~F., \& {Brodie}, J.~P. 2009, \mnras, 400, 2135

\bibitem[{{Foster} {et~al.}(2011){Foster}, {Spitler}, {Romanowsky}, {Forbes},
  {Pota}, {Bekki}, {Strader}, {Proctor}, {Arnold}, \& {Brodie}}]{Foster11}
{Foster}, C., {Spitler}, L.~R., {Romanowsky}, A.~J., {Forbes}, D.~A., {Pota},
  V., {Bekki}, K., {Strader}, J., {Proctor}, R.~N., {Arnold}, J.~A., \&
  {Brodie}, J.~P. 2011, \mnras, 415, 3393

\bibitem[{Furrer {et~al.}(2009)Furrer, Nychka, Sain, \& Nychka}]{Furrer09}
Furrer, R., Nychka, D., Sain, S., \& Nychka, M.~D. 2009, Title Tools for
  spatial data

\bibitem[{{Gavazzi} {et~al.}(2007){Gavazzi}, {Treu}, {Rhodes}, {Koopmans},
  {Bolton}, {Burles}, {Massey}, \& {Moustakas}}]{Gavazzi07}
{Gavazzi}, R., {Treu}, T., {Rhodes}, J.~D., {Koopmans}, L.~V.~E., {Bolton},
  A.~S., {Burles}, S., {Massey}, R.~J., \& {Moustakas}, L.~A. 2007, \apj, 667,
  176

\bibitem[{{Gerhard} {et~al.}(2001){Gerhard}, {Kronawitter}, {Saglia}, \&
  {Bender}}]{Gerhard01}
{Gerhard}, O., {Kronawitter}, A., {Saglia}, R.~P., \& {Bender}, R. 2001, \aj,
  121, 1936

\bibitem[{{Gerhard}(1993)}]{Gerhard93}
{Gerhard}, O.~E. 1993, \mnras, 265, 213

\bibitem[{{Greene} {et~al.}(2013){Greene}, {Murphy}, {Graves}, {Gunn},
  {Raskutti}, {Comerford}, \& {Gebhardt}}]{Greene13}
{Greene}, J.~E., {Murphy}, J.~D., {Graves}, G.~J., {Gunn}, J.~E., {Raskutti},
  S., {Comerford}, J.~M., \& {Gebhardt}, K. 2013, \apj, 776, 64

\bibitem[{{Hasting}(1970)}]{Hasting70}
{Hasting}, W.~K. 1970, Biometrika, 57, 97

\bibitem[{{Humphrey} \& {Buote}(2010)}]{Humphrey10}
{Humphrey}, P.~J. \& {Buote}, D.~A. 2010, \mnras, 403, 2143

\bibitem[{{Kuntschner} {et~al.}(2010){Kuntschner}, {Emsellem}, {Bacon},
  {Cappellari}, {Davies}, {de Zeeuw}, {Falc{\'o}n-Barroso}, {Krajnovi{\'c}},
  {McDermid}, {Peletier}, {Sarzi}, {Shapiro}, {van den Bosch}, \& {van de
  Ven}}]{Kuntschner10}
{Kuntschner}, H., {Emsellem}, E., {Bacon}, R., {Cappellari}, M., {Davies},
  R.~L., {de Zeeuw}, P.~T., {Falc{\'o}n-Barroso}, J., {Krajnovi{\'c}}, D.,
  {McDermid}, R.~M., {Peletier}, R.~F., {Sarzi}, M., {Shapiro}, K.~L., {van den
  Bosch}, R.~C.~E., \& {van de Ven}, G. 2010, \mnras, 408, 97

\bibitem[{{Ma} {et~al.}(2014){Ma}, {Greene}, {McConnell}, {Janish},
  {Blakeslee}, {Thomas}, \& {Murphy}}]{Ma14}
{Ma}, C.-P., {Greene}, J.~E., {McConnell}, N., {Janish}, R., {Blakeslee},
  J.~P., {Thomas}, J., \& {Murphy}, J.~D. 2014, \apj, 795, 158

\bibitem[{{McDermid} {et~al.}(2015){McDermid}, {Alatalo}, {Blitz}, {Bournaud},
  {Bureau}, {Cappellari}, {Crocker}, {Davies}, {Davis}, {de Zeeuw}, {Duc},
  {Emsellem}, {Khochfar}, {Krajnovi{\'c}}, {Kuntschner}, {Morganti}, {Naab},
  {Oosterloo}, {Sarzi}, {Scott}, {Serra}, {Weijmans}, \& {Young}}]{McDermid15}
{McDermid}, R.~M., {Alatalo}, K., {Blitz}, L., {Bournaud}, F., {Bureau}, M.,
  {Cappellari}, M., {Crocker}, A.~F., {Davies}, R.~L., {Davis}, T.~A., {de
  Zeeuw}, P.~T., {Duc}, P.-A., {Emsellem}, E., {Khochfar}, S., {Krajnovi{\'c}},
  D., {Kuntschner}, H., {Morganti}, R., {Naab}, T., {Oosterloo}, T., {Sarzi},
  M., {Scott}, N., {Serra}, P., {Weijmans}, A.-M., \& {Young}, L.~M. 2015,
  \mnras, 448, 3484

\bibitem[{{Morganti} {et~al.}(2006){Morganti}, {de Zeeuw}, {Oosterloo},
  {McDermid}, {Krajnovi{\'c}}, {Cappellari}, {Kenn}, {Weijmans}, \&
  {Sarzi}}]{Morganti06}
{Morganti}, R., {de Zeeuw}, P.~T., {Oosterloo}, T.~A., {McDermid}, R.~M.,
  {Krajnovi{\'c}}, D., {Cappellari}, M., {Kenn}, F., {Weijmans}, A., \&
  {Sarzi}, M. 2006, \mnras, 371, 157

\bibitem[{{Navarro} {et~al.}(1996){Navarro}, {Frenk}, \& {White}}]{Navarro96}
{Navarro}, J.~F., {Frenk}, C.~S., \& {White}, S.~D.~M. 1996, \apj, 462, 563

\bibitem[{{Newman} {et~al.}(2013){Newman}, {Cooper}, {Davis}, {Faber}, {Coil},
  {Guhathakurta}, {Koo}, {Phillips}, {Conroy}, {Dutton}, {Finkbeiner}, {Gerke},
  {Rosario}, {Weiner}, {Willmer}, {Yan}, {Harker}, {Kassin}, {Konidaris},
  {Lai}, {Madgwick}, {Noeske}, {Wirth}, {Connolly}, {Kaiser}, {Kirby},
  {Lemaux}, {Lin}, {Lotz}, {Luppino}, {Marinoni}, {Matthews}, {Metevier}, \&
  {Schiavon}}]{Newman13}
{Newman}, J.~A., {Cooper}, M.~C., {Davis}, M., {Faber}, S.~M., {Coil}, A.~L.,
  {Guhathakurta}, P., {Koo}, D.~C., {Phillips}, A.~C., {Conroy}, C., {Dutton},
  A.~A., {Finkbeiner}, D.~P., {Gerke}, B.~F., {Rosario}, D.~J., {Weiner},
  B.~J., {Willmer}, C.~N.~A., {Yan}, R., {Harker}, J.~J., {Kassin}, S.~A.,
  {Konidaris}, N.~P., {Lai}, K., {Madgwick}, D.~S., {Noeske}, K.~G., {Wirth},
  G.~D., {Connolly}, A.~J., {Kaiser}, N., {Kirby}, E.~N., {Lemaux}, B.~C.,
  {Lin}, L., {Lotz}, J.~M., {Luppino}, G.~A., {Marinoni}, C., {Matthews},
  D.~J., {Metevier}, A., \& {Schiavon}, R.~P. 2013, \apjs, 208, 5

\bibitem[{{Noordermeer} {et~al.}(2008){Noordermeer}, {Merrifield}, {Coccato},
  {Arnaboldi}, {Capaccioli}, {Douglas}, {Freeman}, {Gerhard}, {Kuijken}, {de
  Lorenzi}, {Napolitano}, \& {Romanowsky}}]{Noordermeer08}
{Noordermeer}, E., {Merrifield}, M.~R., {Coccato}, L., {Arnaboldi}, M.,
  {Capaccioli}, M., {Douglas}, N.~G., {Freeman}, K.~C., {Gerhard}, O.,
  {Kuijken}, K., {de Lorenzi}, F., {Napolitano}, N.~R., \& {Romanowsky}, A.~J.
  2008, \mnras, 384, 943

\bibitem[{{Norris} {et~al.}(2008){Norris}, {Sharples}, {Bridges}, {Gebhardt},
  {Forbes}, {Proctor}, {Faifer}, {Forte}, {Beasley}, {Zepf}, \&
  {Hanes}}]{Norris08}
{Norris}, M.~A., {Sharples}, R.~M., {Bridges}, T., {Gebhardt}, K., {Forbes},
  D.~A., {Proctor}, R., {Faifer}, F.~R., {Forte}, J.~C., {Beasley}, M.~A.,
  {Zepf}, S.~E., \& {Hanes}, D.~A. 2008, \mnras, 385, 40

\bibitem[{{Pastorello} {et~al.}(2014){Pastorello}, {Forbes}, {Foster},
  {Brodie}, {Usher}, {Romanowsky}, {Strader}, \& {Arnold}}]{Pastorello14}
{Pastorello}, N., {Forbes}, D.~A., {Foster}, C., {Brodie}, J.~P., {Usher}, C.,
  {Romanowsky}, A.~J., {Strader}, J., \& {Arnold}, J.~A. 2014, \mnras, 442,
  1003

\bibitem[{{Proctor} {et~al.}(2009){Proctor}, {Forbes}, {Romanowsky}, {Brodie},
  {Strader}, {Spolaor}, {Mendel}, \& {Spitler}}]{Proctor09}
{Proctor}, R.~N., {Forbes}, D.~A., {Romanowsky}, A.~J., {Brodie}, J.~P.,
  {Strader}, J., {Spolaor}, M., {Mendel}, J.~T., \& {Spitler}, L. 2009, \mnras,
  398, 91

\bibitem[{{Raskutti} {et~al.}(2014){Raskutti}, {Greene}, \&
  {Murphy}}]{Raskutti14}
{Raskutti}, S., {Greene}, J.~E., \& {Murphy}, J.~D. 2014, \apj, 786, 23

\bibitem[{{Saglia} {et~al.}(2010){Saglia}, {Fabricius}, {Bender}, {Montalto},
  {Lee}, {Riffeser}, {Seitz}, {Morganti}, {Gerhard}, \& {Hopp}}]{Saglia10}
{Saglia}, R.~P., {Fabricius}, M., {Bender}, R., {Montalto}, M., {Lee}, C.-H.,
  {Riffeser}, A., {Seitz}, S., {Morganti}, L., {Gerhard}, O., \& {Hopp}, U.
  2010, \aap, 509, A61

\bibitem[{{Salpeter}(1955)}]{Salpeter55}
{Salpeter}, E.~E. 1955, \apj, 121, 161

\bibitem[{{S{\'a}nchez} {et~al.}(2012){S{\'a}nchez}, {Kennicutt}, {Gil de Paz},
  {van de Ven}, {V{\'{\i}}lchez}, {Wisotzki}, {Walcher}, {Mast}, {Aguerri},
  {Albiol-P{\'e}rez}, {Alonso-Herrero}, {Alves}, {Bakos}, {Bart{\'a}kov{\'a}},
  {Bland-Hawthorn}, {Boselli}, {Bomans}, {Castillo-Morales}, {Cortijo-Ferrero},
  {de Lorenzo-C{\'a}ceres}, {Del Olmo}, {Dettmar}, {D{\'{\i}}az}, {Ellis},
  {Falc{\'o}n-Barroso}, {Flores}, {Gallazzi}, {Garc{\'{\i}}a-Lorenzo},
  {Gonz{\'a}lez Delgado}, {Gruel}, {Haines}, {Hao}, {Husemann},
  {Igl{\'e}sias-P{\'a}ramo}, {Jahnke}, {Johnson}, {Jungwiert}, {Kalinova},
  {Kehrig}, {Kupko}, {L{\'o}pez-S{\'a}nchez}, {Lyubenova}, {Marino},
  {M{\'a}rmol-Queralt{\'o}}, {M{\'a}rquez}, {Masegosa}, {Meidt},
  {Mendez-Abreu}, {Monreal-Ibero}, {Montijo}, {Mour{\~a}o}, {Palacios-Navarro},
  {Papaderos}, {Pasquali}, {Peletier}, {P{\'e}rez}, {P{\'e}rez}, {Quirrenbach},
  {Rela{\~n}o}, {Rosales-Ortega}, {Roth}, {Ruiz-Lara},
  {S{\'a}nchez-Bl{\'a}zquez}, {Sengupta}, {Singh}, {Stanishev}, {Trager},
  {Vazdekis}, {Viironen}, {Wild}, {Zibetti}, \& {Ziegler}}]{Sanchez12}
{S{\'a}nchez}, S.~F., {Kennicutt}, R.~C., {Gil de Paz}, A., {van de Ven}, G.,
  {V{\'{\i}}lchez}, J.~M., {Wisotzki}, L., {Walcher}, C.~J., {Mast}, D.,
  {Aguerri}, J.~A.~L., {Albiol-P{\'e}rez}, S., {Alonso-Herrero}, A., {Alves},
  J., {Bakos}, J., {Bart{\'a}kov{\'a}}, T., {Bland-Hawthorn}, J., {Boselli},
  A., {Bomans}, D.~J., {Castillo-Morales}, A., {Cortijo-Ferrero}, C., {de
  Lorenzo-C{\'a}ceres}, A., {Del Olmo}, A., {Dettmar}, R.-J., {D{\'{\i}}az},
  A., {Ellis}, S., {Falc{\'o}n-Barroso}, J., {Flores}, H., {Gallazzi}, A.,
  {Garc{\'{\i}}a-Lorenzo}, B., {Gonz{\'a}lez Delgado}, R., {Gruel}, N.,
  {Haines}, T., {Hao}, C., {Husemann}, B., {Igl{\'e}sias-P{\'a}ramo}, J.,
  {Jahnke}, K., {Johnson}, B., {Jungwiert}, B., {Kalinova}, V., {Kehrig}, C.,
  {Kupko}, D., {L{\'o}pez-S{\'a}nchez}, {\'A}.~R., {Lyubenova}, M., {Marino},
  R.~A., {M{\'a}rmol-Queralt{\'o}}, E., {M{\'a}rquez}, I., {Masegosa}, J.,
  {Meidt}, S., {Mendez-Abreu}, J., {Monreal-Ibero}, A., {Montijo}, C.,
  {Mour{\~a}o}, A.~M., {Palacios-Navarro}, G., {Papaderos}, P., {Pasquali}, A.,
  {Peletier}, R., {P{\'e}rez}, E., {P{\'e}rez}, I., {Quirrenbach}, A.,
  {Rela{\~n}o}, M., {Rosales-Ortega}, F.~F., {Roth}, M.~M., {Ruiz-Lara}, T.,
  {S{\'a}nchez-Bl{\'a}zquez}, P., {Sengupta}, C., {Singh}, R., {Stanishev}, V.,
  {Trager}, S.~C., {Vazdekis}, A., {Viironen}, K., {Wild}, V., {Zibetti}, S.,
  \& {Ziegler}, B. 2012, \aap, 538, A8

\bibitem[{{Savorgnan} \& {Graham}(2015)}]{Savorgnan15}
{Savorgnan}, G.~A.~D. \& {Graham}, A.~W. e.~a. 2015, \mnras

\bibitem[{{Scott} {et~al.}(2009){Scott}, {Cappellari}, {Davies}, {Bacon}, {de
  Zeeuw}, {Emsellem}, {Falc{\'o}n-Barroso}, {Krajnovi{\'c}}, {Kuntschner},
  {McDermid}, {Peletier}, {Pipino}, {Sarzi}, {van den Bosch}, {van de Ven}, \&
  {van Scherpenzeel}}]{Scott09}
{Scott}, N., {Cappellari}, M., {Davies}, R.~L., {Bacon}, R., {de Zeeuw}, P.~T.,
  {Emsellem}, E., {Falc{\'o}n-Barroso}, J., {Krajnovi{\'c}}, D., {Kuntschner},
  H., {McDermid}, R.~M., {Peletier}, R.~F., {Pipino}, A., {Sarzi}, M., {van den
  Bosch}, R.~C.~E., {van de Ven}, G., \& {van Scherpenzeel}, E. 2009, \mnras,
  398, 1835

\bibitem[{{Statler}(1994)}]{Statler94}
{Statler}, T.~S. 1994, \aj, 108, 111

\bibitem[{{Statler} \& {Smecker-Hane}(1999)}]{Statler99}
{Statler}, T.~S. \& {Smecker-Hane}, T. 1999, \aj, 117, 839

\bibitem[{{Thomas} {et~al.}(2011){Thomas}, {Saglia}, {Bender}, {Thomas},
  {Gebhardt}, {Magorrian}, {Corsini}, {Wegner}, \& {Seitz}}]{Thomas11}
{Thomas}, J., {Saglia}, R.~P., {Bender}, R., {Thomas}, D., {Gebhardt}, K.,
  {Magorrian}, J., {Corsini}, E.~M., {Wegner}, G., \& {Seitz}, S. 2011, \mnras,
  415, 545

\bibitem[{{Usher} {et~al.}(2012){Usher}, {Forbes}, {Brodie}, {Foster},
  {Spitler}, {Arnold}, {Romanowsky}, {Strader}, \& {Pota}}]{Usher12}
{Usher}, C., {Forbes}, D.~A., {Brodie}, J.~P., {Foster}, C., {Spitler}, L.~R.,
  {Arnold}, J.~A., {Romanowsky}, A.~J., {Strader}, J., \& {Pota}, V. 2012,
  \mnras, 426, 1475

\bibitem[{{van der Marel} \& {Franx}(1993)}]{vanderMarel93}
{van der Marel}, R.~P. \& {Franx}, M. 1993, \apj, 407, 525

\bibitem[{{Vazdekis} {et~al.}(2010){Vazdekis}, {S{\'a}nchez-Bl{\'a}zquez},
  {Falc{\'o}n-Barroso}, {Cenarro}, {Beasley}, {Cardiel}, {Gorgas}, \&
  {Peletier}}]{Vazdekis10}
{Vazdekis}, A., {S{\'a}nchez-Bl{\'a}zquez}, P., {Falc{\'o}n-Barroso}, J.,
  {Cenarro}, A.~J., {Beasley}, M.~A., {Cardiel}, N., {Gorgas}, J., \&
  {Peletier}, R.~F. 2010, \mnras, 404, 1639

\bibitem[{{Weijmans} {et~al.}(2009){Weijmans}, {Cappellari}, {Bacon}, {de
  Zeeuw}, {Emsellem}, {Falc{\'o}n-Barroso}, {Kuntschner}, {McDermid}, {van den
  Bosch}, \& {van de Ven}}]{Weijmans09}
{Weijmans}, A.-M., {Cappellari}, M., {Bacon}, R., {de Zeeuw}, P.~T.,
  {Emsellem}, E., {Falc{\'o}n-Barroso}, J., {Kuntschner}, H., {McDermid},
  R.~M., {van den Bosch}, R.~C.~E., \& {van de Ven}, G. 2009, \mnras, 398, 561

\bibitem[{{White} \& {Rees}(1978)}]{White78b}
{White}, S.~D.~M. \& {Rees}, M.~J. 1978, \mnras, 183, 341

\end{thebibliography}

\end{document}